\documentclass[12pt,draftclsnofoot,onecolumn]{IEEEtran}
\usepackage[T1]{fontenc}% optional T1 font encoding
\usepackage{cite}
\usepackage{amsthm,amsmath,amssymb,bm}
\usepackage{mathrsfs}
\usepackage{graphicx}
\usepackage{enumitem}
\usepackage{algorithm}  
\usepackage{algorithmic}  
\usepackage{array}
\usepackage{subfigure}
\usepackage{psfrag}
\usepackage{booktabs}
\usepackage{tabularx}
\usepackage{xcolor}
\usepackage[caption=false]{subfig}

\ifCLASSINFOpdf
\else
\fi


\begin{document}
\renewcommand{\qedsymbol}{}
%	\title{Analytical Solution of Beamforming and Hierarchical Codebook Design on Intelligent Reflecting Surface}

\title{Hierarchical Codebook Design and Analytical Beamforming Solution for IRS Assisted Communication }

% Robust Hierarchical Codebook Design on Intelligent Reflecting Surfaces: An Analytical Solution for Passive Beamforming
\author{
	Xiyuan~Liu,~\IEEEmembership{Graduate Student Member,~IEEE,}
	Qingqing~Wu,~\IEEEmembership{Senior Member,~IEEE,}
	Die~Hu,~\IEEEmembership{Member,~IEEE,}
	Rui~Wang,~\IEEEmembership{Senior Member,~IEEE,}
	Jun~Wu,~\IEEEmembership{Senior Member,~IEEE,}
	\thanks{X. Liu is with the College of Electronics and Information Engineering, Tongji University, Shanghai 201804, China (e-mail: 1910670@tongji.edu.cn).
		Q. Wu is with the Department of Electronic Engineering, Shanghai Jiao Tong University, 200240, China (e-mail: qingqingwu@sjtu.edu.cn).
		D. Hu is with the Key Laboratory of EMW Information, Fudan University, Shanghai 200433, China (e-mail: hudie@fudan.edu.cn).
		R. Wang is with the College of Electronics and Information Engineering, Tongji University, Shanghai 201804, China, and also with the Shanghai Institute of Intelligent Science and Technology, Tongji University, Shanghai 201804, China (e-mail: ruiwang@tongji.edu.cn).
		J. Wu is with the School of Computer Science, Fudan University, Shanghai 200433, China, and also with the Shanghai Qi Zhi Institute, Shanghai 200030, China (e-mail: wujun@fudan.edu.cn).}	
}	

\maketitle




% for Transactions on Magnetics papers, we must declare the abstract and
% index terms PRIOR to the title within the \IEEEtitleabstractindextext
% IEEEtran command as these need to go into the title area created by
\IEEEtitleabstractindextext{%
	\begin{abstract}
		
		%			In intelligent reflecting surface(IRS) aided communication scenario, beam searching is always  time consuming, this is because IRS needs to serve both incident and reflect channel, so it needs double time to search. However, incident channel and reflect channel can be combined as cascaded channel. This paper propose an IRS hierarchical codebook algorithm based on cascaded channel. The definition of cascaded channel direction is proposed firstly and then a new cascaded channel array factor is put forward. The difficulty of hierarchical codebook design is how to form wide beams. Analytical solution of IRS beamforming problem is given in this paper. Based on the analytical solution we put forward linear phase shift difference (LPD) algorithm to form wide beams. We put forward two IRS hierarchical codebooks based on LPD algorithm. Simulation results show LPD codebook can control beam direction, beam width and beam power distribution precisely, and it can decouple these three variables. The time complexity of LPD algorithm is o(1), which is irrelevant to array size. So LPD algorithm can get results in a short time even on large size arrays. LPD algorithm  can realize same funtion on antenna array of different size or shape, it is the first beamforming algorithm which  robust to array size. When the array size approach infinity LPD beam shows Gibbs phenomenon. This paper is the first work which find Gibbs phenomenon on beamforming area. 
		%		
		In intelligent reflecting surface (IRS) assisted communication, beam search is usually time-consuming as the multiple-input multiple-output  (MIMO) of IRS is usually very large. Hierarchical codebooks is a widely accepted method for reducing the complexity of searching time. The performance of this method strongly depends on the design scheme of beamforming of different beamwidths. In this paper, a non-constant phase difference (NCPD) beamforming algorithm  is proposed. To implement  the NCPD algorithm, we first model the phase shift of IRS as a  continuous function, and then determine  the parameters of the continuous function through the analysis of its array factor. Then, we propose a hierarchical codebook and two beam training schemes, namely the joint searching (JS) scheme  and direction-wise searching (DWS) scheme by using the NCPD algorithm which can flexibly change the width, direction and shape of the beam formed by the IRS array.  Simulation results show that the NCPD algorithm is more accurate with smaller side lobes, and also  more stable on IRS of different sizes compared to other wide beam algorithms. The misalignment  rate of the beam formed by the NCPD method  is significantly reduced. The time complexity of the NCPD algorithm  is constant, thus making it  more suitable for solving the beamforming design problem with practically large IRS.

	\end{abstract}
	
	}


\IEEEdisplaynontitleabstractindextext

% \fi
\IEEEpeerreviewmaketitle

\section{Introduction}

Intelligent reflecting surface (IRS) is a key technology often mentioned in 6th generation (6G) white papers. An IRS is generally composed of a large number of passive elements each being able to reflect the incident signal with an adjustable phase shift. Since IRS is passive, its cost is low and large-scale deployment can be achieved within a limited cost. IRS can be deployed on a flat surface such as the outer surface of a building to enhance the signal quality in a blind area of a base station  \cite{DBLP:journals/tcom/WuZZYZ21}. Since IRS has no RF chain, the array scale on IRS is usually very large to provide sufficient beamforming gain \cite{DBLP:journals/tcom/WuZZYZ21}. The larger the array scale, the higher is the resolution of the array and the narrower is the formed beam \cite{DBLP:journals/icl/JamaliNSP21}. The narrower the beam, the greater is the gain of the beam, and the better is the effect for communication. But the narrower the beam, the greater is the number of beams, which makes it to take longer time to search for a suitable beam \cite{DBLP:journals/icl/JamaliNSP21}. This type of narrow beams are called as the  pencil beam \cite{DBLP:conf/sigcomm/HassaniehARAKI18}. Therefore, how to speed up the beam search process is an important issue in the IRS-assisted communication scenarios.

	Although the beam search is still immature in the IRS-assisted communication system, it is widely used in the conventional base station (BS) and user equipment (UE) scenarios.  According to whether channel state information (CSI) information is utilized, beam search work can be divided into two categories: CSI based and CSI free. There are four main CSI based algorithms. The first one is the context-based beam search method \cite{DBLP:journals/cm/GiordaniMZ16,DBLP:journals/corr/abs-1711-05456,DBLP:conf/iccchina/LuLDFHG21}, which exploits the user's location or movement related information to assist the  beam search. This method first  needs the user's location information, and then uses the user's position related information  to assist the communication. But sometimes the location information is not accurate enough, which may  cause the selected beam to be misaligned. The second  method is based on machine learning \cite{DBLP:conf/wowmom/DevotiFC18, DBLP:conf/wcnc/ChenCW18, DBLP:journals/access/AlkhateebAVLQT18} where the required beams can be obtained by learning the channel state information (CSI). However, a machine learning method  requires a large amount of data to train the model. The third method uses the  cascaded channel estimation results to optimize the  IRS phase shift matrix. However since IRS can not  supply a continuous phase shift, channel estimation processes  in the  IRS scenario are always not perfect \cite{DBLP:journals/jsac/YouZZ20}. Therefore, some algorithms are made available to study how to optimize the beamforming design of IRS under an inaccurate CSI \cite{DBLP:journals/twc/Papazafeiropoulos22,DBLP:journals/wcl/ZhengZ20,DBLP:journals/wcl/ZhiPRW21a,DBLP:journals/icl/Zhang0Z020,DBLP:journals/twc/ZhengYZ20}. The fourth  method is an optimization algorithm based on the  received signals \cite{DBLP:journals/tcom/WangLJYW21, DBLP:journals/icl/OmidSPDN22,DBLP:journals/jsac/YanYHK20}, such as iterative optimization algorithm \cite{DBLP:journals/twc/PerovicTRF21}.  Because IRS is passive and it is difficult to obtain CSI, CSI based algorithms are not suitable for IRS assisted scenarios. There are three CSI-free methods which are  exhaustive search, hierarchical search and heuristic algorithm, respectively. Exhaustive search method \cite{DBLP:conf/wcnc/WeiLW17} first uses all the beams present in a sequence, then observes the power of the signal obtained by using each beam at the receiving end, and finally selects the corresponding  beam with the highest receiving power. The disadvantage of exhaustive search method is that its search time is too long. Hierarchical method first searches with wide beams to determine the general direction, and then narrows down the search range with narrower beams, and finally identifies the most suitable pencil beam \cite{DBLP:journals/twc/XiaoHXX16,DBLP:journals/wcl/MeiZ21}. The advantage of the hierarchical  method is that it  speeds up the beam search process, but the disadvantage is that the beamforming gain of the wide beams are small, and it is prone to misalignment.
	Heuristic algorithm \cite{DBLP:conf/pimrc/GuoMS17, DBLP:journals/tvt/GaoDYW16, DBLP:conf/ncc/JainRKS22}, such as the  genetic algorithm \cite{DBLP:conf/wiopt/GuoMS17, DBLP:journals/wicomm/GuoMS18},  treats each beam as a gene, and then uses the principles of mutation, recombination and selection of genes to  gradually find the optimal beam. This method also only suit for small arrays since there are too much pencil beams in large arrays scenario which makes it difficult for the algorithm to converge. In this paper, we use the hierarchical algorithm because of its low time complexity, and since we can stop beam searching at any time according to the required beamforming gain and directly use the intermediate results, it is more flexible. In order to implement the hierarchical search algorithm, we need to design a hierarchical codebook in IRS scenarios.

The main difficulty in the design of the IRS cascaded  codebook is how to design the beams of different widths on IRS. Since the design principle of the hierarchical codebook on IRS is similar to that on the active array, we can learn from the wide beamforming methods on the active array. There are mainly three conventional wide beamforming methods. The first method is to turn off a portion of the elements. As the smaller the array, the wider is the beam formation \cite{DBLP:journals/twc/XiaoHXX16}. This method has two major disadvantages. The first disadvantage is that the width of the beam cannot be controlled flexibly, and only a few specific widths can be used. Compared with a narrow beam, a wide beam  loses not only the beamforming gain, but also some additional gain due to the reduction in the array size. The lose in such a gain can be effectively compensated  in active arrays by power amplifier, but it cannot be effectively compensated in IRS. Therefore, this method is not easy to be applied  in the IRS scenario. The second wide beamforming method is the  beam combination \cite{DBLP:conf/icc/Zhang022a}. In this method,  the array is first divided into multiple areas, and then each area forms a sub-beam, and these sub-beams can be used to combine the final desired wide beam. However, the number of divided areas needs to be manually selected according to the array size and targeted beams. So, this algorithm can  be used only if  the array size is known. When the array size changes, the beam combination algorithm needs to be redesigned. In the third method, nonlinear function is used to form IRS phase shift matrix, e.g., a quadratic function \cite{DBLP:journals/icl/JamaliNSP21}. This method can form a more flexible wide beam. In order to address the above issues, in this study  we propose to relax the IRS beamforming problem into a continuous function analysis problem, and then select the parameters of this continuous function to achieve a hierarchical codebook that can be used in arrays of different size. Another typical quadratic function used in wide beamforming is Zadoff-Chu sequence \cite{DBLP:journals/tit/Chu72,DBLP:journals/icl/PengT18}. Zadoff-Chu sequence is not suitable for small arrays, but it performs good on large arrays. However, Zadoff-Chu sequence can only form flat beams, it can not used to form more complicated beams.

The main contributions made in this paper are summarized as below:	

	\begin{itemize}
		\item  According to the characteristics of the cascaded channel via  IRS, we propose a definition of the direction of the cascaded channel. Compared with the conventional form where the elevation and azimuth angles of incident and reflecting paths are used to describe the channel direction, our scheme uses fewer parameters to achieve the same purpose, in which only two parameters are used to replace the original four parameters. In addition, the values of the parameters of the cascaded channel direction defined  in our scheme have a one-to-one mapping with the IRS phase shift configuration scheme. Therefore, the hierarchical search in the proposed cascaded channel direction reduces the time complexity.
		\item For the cascaded channel, we propose a transformed array factor  defined in the direction of the cascaded channel  mentioned above. Using the array factor of the cascaded channel, we can analyze the directions of both incident and reflected signals at IRS at the same time. In order to develop an IRS beamforming scheme  applicable to arrays of any size, we transform the phase shift function of IRS into a continuous function. Then, the array factor is given after applying the continuous phase shift function. Using the beamforming scheme designed based on such an array factor, we need only to select different numbers of points on the continuous phase shift function to obtain the beamforming scheme on an array of the corresponding size. This avoids redesigning different phase shift configurations for arrays of different sizes. 
		\item For designing the continuous phase shift function, we find the analytical solution of the IRS beamforming problem by using the method of functional analysis. Using this analytical solution, we can find the IRS phase shift configuration for any target beam allowed by the array resolution, which reduces the time complexity of getting the phase shift function to $\mathcal{O}(1)$. For the wide beam required in the hierarchical codebook scenario, we propose a non-constant phase shift difference (NCPD) algorithm by applying the above analytical solution of IRS beamforming. Compared with the conventional algorithm that uses the  beam combination method to realize the required wide beam, the NCPD algorithm is more flexible and the power distribution in the wide beam is more uniform. When the number of IRS array elements tends to be infinite, the array factor of the beam formed by the NCPD algorithm converges to the array factor of the target beam, which avoids the problem of deep depression of the beam  in the beam combination methods. Using the wide beam formed by the NCPD algorithm, we implement a hierarchical codebook over one parameter in the direction of the cascaded channel. 
		\item For the problem of having two parameters in the cascaded channel direction, we propose two beam training schemes, namely the  joint searching (JS) scheme and direction-wise  searching (DWS) scheme. The JS scheme searches for two parameters at the same time while the DWS scheme  searches for one parameter first, and then  for the other parameter. Applying the above two beam training schemes, we realize the hierarchical search of the cascaded channel direction on IRS.
		\item Simulation results verify our theoretical findings that the NCPD algorithm is significantly improved compared to the beam combination algorithm on arrays of different sizes and when forming beams of different widths. When the number of array elements increases, we can observe that the power distribution in the middle of the beam is very flat, and the Gibbs phenomenon appears at the edge of the beam, which shows that the accuracy of the NCPD algorithm is far superior to other similar algorithms.

	\end{itemize}

	The remainder of this paper is organized as follows. Section II 	introduces the system and channel models, followed by the proposal of  array factor of IRS in Section III.  Section IV explains the principle of IRS phase shift configuration.  The design of the hierarchical codebook on IRS is proposed in Section V, while in Section VI we show the proformance and time complexity of proposed NCPD algorithm. Simulation results are presented in Section VII. Finally, we conclude the paper
	in Section VIII.
	
	{\itshape Notations}: In this paper, scalars, vectors and matrices are denoted by italic letters, bold-face lower-case and upper-case letters, respectively. The space of ${ x} \times { y}$ complex-valued matrices is denoted by $\mathbb{C}^{{x} \times { y}}$. For a complex-value vector $\bm x$, ${\bm x}\otimes{\bm y}$ denotes the  kronecker product of $\bm x$ and $\bm y$ while  $|{\bm x}|$ denotes its  modulus and diag$({\bm x})$ denotes a diagonal matrix with each diagonal entry being the corresponding entry in $\bm x$.  For a function ${\bm y}=H({\bm x})$, $H^{-1}({\bm y})$ denotes its inverse function. For a general matrix $\bm A$, ${\bm A}^*$,${\bm A}^H$, and ${\bm A}[i,j]$ denote its conjugate, conjugate transpose, and the $(i,j)$th entry, respectively. $\jmath$ denotes the imaginary unit, i.e., ${\jmath}^2=-1$.  
	
	\section{System   and Channel Models   }
	
	%% and "HIS" in caps to complete the first word.
	%\IEEEPARstart{T}{his} demo file is intended to serve as a ``starter file''
	%for IEEE \textsc{Transactions on Magnetics} journal papers produced under \LaTeX\ using
	%IEEEtran.cls version 1.8b and later.
	%% You must have at least 2 lines in the paragraph with the drop letter
	%% (should never be an issue)
	%I wish you the best of success.
	\begin{figure}
		\centering
		\includegraphics[scale=1]{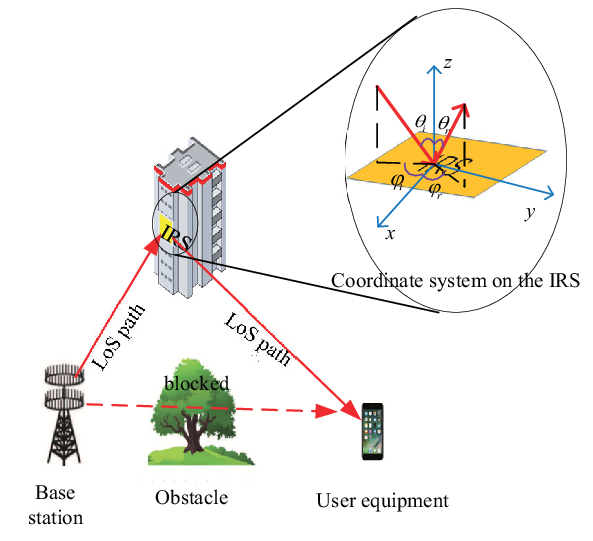}
		\caption{ An IRS assisted communication system.}     \label{cascaded channel}
	\end{figure}
	For the purpose of exposition, we consider a basic IRS-aided  single-user communication system in millimeter-wave (mmWave) band  as illustrated in Fig. \ref{cascaded channel}, where an IRS composed of a large number of $N=N_{\rm ver} \times N_{\rm hor}$ passive reflecting elements, is deployed in proximity to a user for assisting its data transmission with an BS, both of which are equipped with a single antenna. The results in this paper can be readily extended to the more general system with multiple users served by the IRS (e.g., by applying orthogonal time/frequency division multiple access) and/or multiple antennas at the BS (by estimating their associated channels in parallel). In this paper, it assumed that the line of sight (LoS) path of BS-UE channel is blocked while those of BS-IRS and IRS-UE channel are not blocked as shown in Fig. \ref{cascaded channel}. Therefore, the IRS establishes a virtual LoS path of BS-IRS-UE. The received signal of the UE is:

		\begin{equation}\label{received signal}
			y={\bm h}^{H}_{\rm r}{\bm \Phi} {\bm h}_{\rm i}x+z,
		\end{equation}
	
	where $x$, $y$ and $z \sim N{\rm{(}}0{\rm{,}}{\sigma ^2}{\rm{)}}$ are transmitted signal of BS, received signal of UE and channel noise with with zero mean and variance  ${\sigma ^2}$, respectively. ${\bm h}_{\rm r}$ and ${\bm h}_{\rm i}$ are channel of IRS-UE and BS-IRS respectively. $\bm \Phi$ is a diagonal matrix with elements' response value on its diagonal. $\bm \Phi$ can be expressed as:
\begin{equation}\label{Phi}
			{\bm \Phi} = {\rm diag}(e^{\jmath kdg(1)}, e^{\jmath kdg(2)}, \cdots, e^{\jmath kdg(n)},\cdots, e^{\jmath kdg(N)}),
	\end{equation}
	where $n$, $k$ and $d$ are the index of IRS elements, wavenumber and spacing distance of IRS respectively. $k=\frac{2\pi}{\lambda}$ where $\lambda$ is the wave length.  $kdg(n)$ is the phase shift configuration of IRS elements. 
	
	{\itshape Proposition 1}: The maximum spacing between the elements of IRS is $\frac{\lambda}{4}$.
	
	{\itshape Proof}: See Appendix \ref{appendixB}.
	
	In this paper, we choose $d=\frac{\lambda}{4}$. Since mmWave channels are sparse, most of the signal power exists in the LoS paths. So, we ignore the NLoS path and consider only the LoS path. Then ${\bm h}_{\rm r}$ and ${\bm h}_{\rm i}$ can be expressed as:
	\begin{align}
		{\bm h}_{\rm r}=\varrho_{\rm r} {\bm d},\\
		{\bm h}_{\rm i}=\varrho_{\rm i} {\bm a},
	\end{align}
	where $\varrho_{\rm r}$ and $\varrho_{\rm i}$ are path loss of the IRS-UE and BS-IRS channels, respectively. $\bm d$ and $\bm a$ are the steering vectors corresponding  to the angles-of-departure  (AoDs) and angles-of-arrival (AoAs) of IRS, respectively. Therefore, (\ref{received signal}) can be rewritten as:
	\begin{equation}\label{dphia}
		y=\varrho_{\rm r}\varrho_{\rm i}{\bm d}^H{\bm \Phi}{\bm a}x.
	\end{equation}
	Then the change in variables ${\bm d} {\bm \Phi}{\bm a}={\bm \nu}^H {\bm h}_{\rm vtl}$ is applied, where ${\bm \nu}=[e^{jkdg(1)}, e^{jkdg(2)}, \cdots, e^{jkdg(N)}]^T$. ${\bm h}_{\rm vtl}={\rm  diag}({\bm d}){\bm a}$ is an equivalent virtual LoS path of BS-UE. Since IRS is an uniform planar array (UPA), $\bm d$ and $\bm a$ can be disassembled into the following form:
	\begin{align}
		{\bm d}={\bm d}_{\rm hor}\otimes {\bm d}_{\rm ver},\\
		{\bm a}={\bm a}_{\rm hor}\otimes {\bm a}_{\rm ver},
	\end{align}
	where ${\bm d}_{\rm hor}$, ${\bm a}_{\rm hor}$, ${\bm d}_{\rm ver}$ and ${\bm a}_{\rm ver}$ are steering vectors of horizontal and vertical direction of IRS array, which can be expressed as:
		\begin{align}
			&{{\bm d}_{\rm hor}} = [1,{e^{{\jmath}kd\sin {\theta _{\rm r}}\cos {\varphi _{\rm r}}}},{e^{{\jmath}kd2\sin {\theta _{\rm r}}\cos {\varphi _{\rm r}}}}, \cdots {e^{{\jmath}kd(N_{\rm hor} - 1)\sin {\theta _{\rm r}}\cos {\varphi _{\rm r}}}}]^{\mathrm  T},\\
			&{{\bm d}_{\rm ver}} = [1,{e^{{\jmath}kd\sin {\theta _{\rm r}}\sin {\varphi _{\rm r}}}},{e^{{\jmath}kd2\sin {\theta _{\rm r}}\sin {\varphi _{\rm r}}}}, \cdots {e^{{\jmath}kd(N_{\rm ver} - 1)\sin {\theta _{\rm r}}\sin {\varphi _{\rm r}}}}]^{\mathrm  T},\\
			&{{\bm a}_{\rm hor}} = [1,{e^{{\jmath}kd\sin {\theta _{\rm i}}\cos {\varphi _{\rm i}}}},{e^{{\jmath}kd2\sin {\theta _{\rm i}}\cos {\varphi _{\rm i}}}}, \cdots {e^{{\jmath}kd(N_{\rm hor} - 1)\sin {\theta _{\rm i}}\cos {\varphi _{\rm i}}}}]^{\mathrm  T},\\
			&{{\bm a}_{\rm ver}} = [1,{e^{{\jmath}kd\sin {\theta _{\rm i}}\sin {\varphi _{\rm i}}}},{e^{{\jmath}kd2\sin {\theta _{\rm i}}\sin {\varphi _{\rm i}}}}, \cdots {e^{{\jmath}kd(N_{\rm ver} - 1)\sin {\theta _{\rm i}}\sin {\varphi _{\rm i}}}}]^{\mathrm  T},
		\end{align}
		where $\theta_{\rm i}$, $\theta_{\rm r}$, $\phi_{\rm i}$ and $\phi_{\rm r}$ are elevation and azimuth angle of corresponding AoA and AoD of the IRS.
	Hence
	\begin{equation}
		{\bm h}_{\rm vtl}=({\rm diag}({\bm d}_{\rm hor}){\bm a}_{\rm hor})\otimes({\rm diag}({\bm d}_{\rm ver}){\bm a}_{\rm ver}).
	\end{equation}
	We define
	\begin{align}
		&{\bm h}_{\rm hor}={\rm diag}({\bm d}_{\rm hor}){\bm a}_{\rm hor},\\
		&{\bm h}_{\rm ver}={\rm diag}({\bm d}_{\rm ver}){\bm a}_{\rm ver}.
	\end{align}
	Thus ${\bm h}_{\rm hor}$ and ${\bm h}_{\rm ver}$ can be expressed as:
	\begin{align}
		&{\bm h}_{\rm hor}=[1, e^{\jmath kd\beta_{\rm hor}}, e^{\jmath kd2\beta_{\rm hor}}, \cdots, e^{\jmath kd(N_{\rm hor}-1)\beta_{\rm hor}}]^T,\\
		&{\bm h}_{\rm ver}=[1, e^{\jmath kd\beta_{\rm ver}}, e^{\jmath kd2\beta_{\rm ver}}, \cdots, e^{\jmath kd(N_{\rm hor}-1)\beta_{\rm ver}}]^T,	
	\end{align}
	where $\beta_{\rm hor}$ and $\beta_{\rm ver}$ can be written as \cite{DBLP:journals/wcl/CaiYYHLZ21}:
	\begin{align}
		&{\beta _{\rm hor}} = \sin {\theta _{\rm r}}\cos {\varphi _{\rm r}} + \sin {\theta _{\rm i}}\cos {\varphi _{\rm i}},\label{betah}\\
		&{\beta _{\rm ver}} = \sin {\theta _{\rm r}}\sin {\varphi _{\rm r}} + \sin {\theta _{\rm i}}\sin {\varphi _{\rm i}}.\label{betav}	
	\end{align}
	From above we can see that the direction of ${\bm h}_{\rm vtl}$ can be determined by direction of ${\bm h}_{\rm hor}$ and ${\bm h}_{\rm ver}$ which are controlled by $\beta_{\rm hor}$ and $\beta_{\rm ver}$. Simultaneously, $\beta_{\rm hor}$ and $\beta_{\rm ver}$ can be determined by $\theta_{\rm r}$, $\theta_{\rm i}$, $\phi_{\rm r}$ and $\phi_{\rm i}$. Thus we define $\beta_{\rm hor}$ and $\beta_{\rm ver}$ are cascaded channel angles of the direction ${\bm h}_{\rm hor}$ and ${\bm h}_{\rm ver}$, respectively. The following sections will use $\beta_{\rm hor}$ and $\beta_{\rm ver}$ to describe cascaded channel direction rather than $\theta_{\rm r}$, $\theta_{\rm i}$, $\phi_{\rm r}$ and $\phi_{\rm i}$.
	
	\section{IRS Array Factor}
	In order to realize the hierarchical codebook on IRS, we need to design an algorithm that can flexibly control the orientation, width and shape of beam on the IRS.  In this section we will introduce array factor of IRS and then analysis beamforming goal of hierarchical codebook on the proposed continuous array factor.
	
	According to the array factor of a conventional array \cite{DBLP:conf/wcnc/WeiLW17}, the array factor modulus (AFM)  on the IRS can be expressed as:
	\begin{equation}\label{array factor}
		\begin{array}{l}
			AFM(\beta_{\rm hor},\beta_{\rm ver})=|{\bm d}^H{\bm \Phi}{\bm a}|\\
			=|{\bm \nu}^H {\bm h}_{\rm vtl}|\\
			= \Bigg|\sum\limits_{{n_{\rm hor}} = 0}^{N_{\rm hor} - 1} {\sum\limits_{{n_{\rm ver}} = 0}^{N_{\rm ver} - 1} {{e^{{\jmath}kd[{\beta _{\rm hor}}{n_{\rm hor}} + {\beta _{\rm ver}}{n_{\rm ver}} + g[(n_{\rm hor}-1)N_{\rm hor}+n_{\rm ver}]]}}} } \Bigg|.\\
		\end{array}	
	\end{equation}
	Since $\beta_{\rm hor}n_{\rm hor}$ and $\beta_{\rm ver}n_{\rm ver}$ are independent to each other, so we can  restrict $ g[(n_{\rm hor}-1)N_{\rm hor}+n_{\rm ver}]$ to the form in which $g[(n_{\rm hor}-1)N_{\rm hor}+n_{\rm ver}]$ is  the sum of two independent parts as below:
	\begin{equation}\label{guv}
		g[(n_{\rm hor}-1)N_{\rm hor}+n_{\rm ver}]=g_{\rm hor}(n_{\rm hor})+g_{\rm ver}(n_{\rm ver}),	
	\end{equation}
	where  $n_{\rm hor}$ and $n_{\rm ver}$ are elements' index along horizontal and vertical on IRS array  and 
	$g_{\rm hor}(n_{\rm hor})$ and $g_{\rm ver}(n_{\rm ver})$ are the  IRS phase shift functions along the  horizontal  and vertical directions on IRS array, respectively. Thus (\ref{array factor}) can be rewritten as:
	\begin{equation}\label{array factor final}
		\begin{array}{l}
			AFM({\beta _{\rm hor}},{\beta _{\rm ver}}) 
			= \Bigg|\sum\limits_{{n_{\rm hor}} = 0}^{N_{\rm hor} - 1} {\sum\limits_{{n_{\rm ver}} = 0}^{N_{\rm ver} - 1} {{e^{{\jmath}kd[{\beta _{\rm hor}}{n_{\rm hor}} + {\beta _{\rm ver}}{n_{\rm ver}} + {g_{\rm hor}}({n_{\rm hor}}) + {g_{\rm ver}}({n_{\rm ver}})]}}} } \Bigg|\\
			= \Bigg|\sum\limits_{{n_{\rm hor}} = 0}^{N_{\rm hor} - 1} {{e^{{\jmath}kd[{\beta _{\rm hor}}{n_{\rm hor}} + {g_{\rm hor}}({n_{\rm hor}})]}}} \sum\limits_{{n_{\rm ver}} = 0}^{N_{\rm ver} - 1} {{e^{{\jmath}kd[{\beta _{\rm ver}}{n_{\rm ver}} + {g_{\rm ver}}({n_{\rm ver}})]}}} \Bigg|\\
			= \Bigg|\sum\limits_{{n_{\rm hor}} = 0}^{N_{\rm hor} - 1} {{e^{{\jmath}kd[{\beta _{\rm hor}}{n_{\rm hor}} + {g_{\rm hor}}({n_{\rm hor}})]}}} \Bigg|\cdot\Bigg|\sum\limits_{{n_{\rm ver}} = 0}^{N_{\rm ver} - 1} {{e^{{\jmath}kd[{\beta _{\rm ver}}{n_{\rm ver}} + {g_{\rm ver}}({n_{\rm ver}})]}}} \Bigg|.
		\end{array}
	\end{equation}
	We can see in  (\ref{array factor final}) that the IRS reflected power is the product of the reflected powers in the ${\bm h}_{\rm hor}$  and  ${\bm h}_{\rm ver}$ directions. The reflected powers of  IRS along the ${\bm h}_{\rm hor}$ and ${\bm h}_{\rm ver}$ directions are two independent parts, and they have duality. So we  take only the ${\bm h}_{\rm hor}$ direction as an example to explain how to determine the optimal $g_{\rm hor}(n_{\rm hor})$, and then the ${\bm h}_{\rm ver}$ direction can be obtained in the same way. The array factor on ${\bm h}_{\rm hor}$ direction can be expressed as:
	\begin{equation}\label{u array factor}
		AFM({\beta _{\rm hor}}) = \Bigg|\sum\limits_{{n_{\rm hor}} = 0}^{N_{\rm hor} - 1} {{e^{{\jmath}kd[{\beta _{\rm hor}}{n_{\rm hor}} + {g_{\rm hor}}({n_{\rm hor}})]}}} \Bigg|.
	\end{equation}
	We normalize (\ref{u array factor}) to get the following:
	\begin{equation}\label{normed u array factor}
		AFM_{\rm norm}({\beta _{\rm hor}}) = \frac{1}{N_{\rm hor}}\Bigg|\sum\limits_{{n_{\rm hor}} = 0}^{N_{\rm hor} - 1} {{e^{{\jmath}kd[{\beta _{\rm hor}}{n_{\rm hor}} + {g_{\rm hor}}({n_{\rm hor}})]}}} \Bigg|.
	\end{equation}
	
	{\itshape Proposition 2}: As $d \to 0$, we have:
	\begin{equation}\label{g_represent}
		AFM_{\rm norm}({\beta _{\rm hor}}) = \Bigg|\int_0^1 {{e^{{\jmath}kN_{\rm hor}d[{\beta _{\rm hor}}\ddot{u}  - \int_0^{\ddot{u}} {\dot{f}(\mu){\rm d}{\mu}} ]}}} {\rm d}{\ddot{u}} \Bigg|.
	\end{equation}
	
	{\itshape Proof}: See Appendix \ref{appendixA}.
	
	Next, we first get the corresponding ${\dot{f}(\mu)}$ according to the goal of beamforming, and then  get the final $g_{\rm hor}(n_{\rm hor})$.

	The direction of the narrow beam is a certain cascaded channel direction. If  the targeted  cascaded channel direction is $\psi_{{\rm hor},0}$, the goal of the  narrow beam can be expressed as:
	\begin{equation}\label{narrow beam}
		AFM_{\rm norm}({\beta _{\rm hor}}) = \left\{ \begin{array}{l}
			1,\,\;\;{\rm if}\;{\beta _{\rm hor}} = {\psi _{{\rm hor},0}},\\
			0,\;\;\;{\rm otherwise}.
		\end{array} \right.
	\end{equation}
	The maximum of AFM is $1$ for the reason that:
	\begin{equation}\label{lessequal}
		\Bigg|\int_0^1 {{e^{{\jmath}kN_{\rm hor}d[{\beta _{\rm hor}}\ddot{u}  - \int_0^{\ddot{u}} {\dot{f}(\mu){\rm d}{\mu}} ]}}} {\rm d}{\ddot{u}} \Bigg| \leq \int_0^1 \Bigg|{{e^{{\jmath}kN_{\rm hor}d[{\beta _{\rm hor}}\ddot{u}  - \int_0^{\ddot{u}} {\dot{f}(\mu){\rm d}{\mu}} ]}}} \Bigg|{\rm d}{\ddot{u}}  \leq 1 .
	\end{equation}
	When $\dot{f}(\mu)=\psi_{{\rm hor},0}$, $AFM_{\rm norm}({\psi_{{\rm hor},0}})=1$. So one of the solution of (\ref{narrow beam}) is $\dot{f}(\mu)=\psi_{{\rm hor},0}$.
	
	The wide beam refers to the beam  oriented towards a range of the cascaded channel directions. Suppose that  cascaded channel direction range of the target wide beam is $(\psi_{{\rm hor},a},\psi_{{\rm hor},b}) $, and the amplitude  distribution of the wide beam is $h(\beta_{\rm hor})$. Then, the wide beam array factor goal can be expressed as:
	\begin{equation}\label{widebeamcondition}
		AFM_{\rm norm}({\beta _{\rm hor}}) = \left\{ \begin{array}{l}
			h(\beta_{\rm hor}),\;\;\;\;\;\;\;{\beta _{\rm hor}} \in [{\psi _{{\rm hor},a}},{\psi _{{\rm hor},b}}],\\
			0,\;\;\;\;\;\;\;\;\;\;{\beta _{\rm hor}} \notin [{\psi _{{\rm hor},a}},{\psi _{{\rm hor},b}}].
		\end{array} \right.
	\end{equation}
	The following sections will elaborate how to design $\dot{f}(\mu)$ to satisfy this wide beam requirement and how to use wide beams to form hierarchical codebook on IRS.

	%The resolution of IRS depends on the size of the IRS \cite{DBLP:journals/jsac/SohrabiCY21}. Hence for any given array size, there is a corresponding narrowest beamwidth. We call the beam that reaches the narrowest width on  the given array size as the narrow beam, and others are called as the wide beam. The width of the narrow beam is determined by the array resolution, rather than by a defined width.\cite{DBLP:journals/tvt/NingCCDF21}.

	\section{Configuration of IRS Phase Shift for Beamforming of Wide Beam}
		In this section, we will discuss how to design $\dot{f}(\mu)$ by satisfying the goal of wide beam as in (\ref{widebeamcondition}).
		In order to  simplify the problem, we specify that $\dot{f}(\mu)$ is an  increasing differentiable function in the range of  $[\dot{f}(0), \dot{f}(1)]$. 
		%	We take an infinitesimal quantity $\delta$ and divide $[\dot{f}(0), \dot{f}(1)]$ into an infinite number of micro-element intervals of width  $\delta$.  Assuming that the micro-element interval at $\mu_0$ is $(\mu_0,\mu_0+\Delta \mu_0)$, $\dot{f}(\mu)$ considered to have  a fixed value of $\dot{f}(\mu_0)$ in this interval.	
		We take a  infinitesimal interval $(\mu_0,\mu_0+\Delta \mu_0)$ and replace $\dot{f}(\mu)$ with its first-order Taylor approximation  on this element: $\dot{f}(\mu)=\dot{f}(\mu_0)+\dot{f}'(\mu_0)(\mu-\mu_0)$.
		
		In the infinitesimal interval $(\mu_0, \mu_0+\Delta \mu_0)$ at $\mu_0$, (\ref{g_represent}) can be written as:
		\begin{equation}\label{microelement on mu}
			\begin{array}{l}
				AFM_{\rm norm}({\beta _{\rm hor}}){|_{\ddot{u}  \in ({\mu _0},{\mu _0} + \Delta \mu_0 )}}\\
				= \Bigg|\int_{{\mu _0}}^{{\mu _0} + \Delta \mu_0 } {{e^{{\jmath}kN_{\rm hor}d[{\beta _{\rm hor}}\ddot{u} - \int_0^{\ddot{u}}  {\dot{f}(\mu_0)+\dot{f}'(\mu_0)(\mu-\mu_0){\rm d}{\mu}} ]}}} {\rm d}{\ddot{u}} \Bigg|\\
				= \Bigg|\int_{{\mu _0}}^{{\mu _0} + \Delta \mu_0 } {{e^{{\jmath}kN_{\rm hor}d[{\beta _{\rm hor}}{\mu _0} - \int_0^{{\mu _0}} {\dot{f}(\mu){\rm d}{\mu}} ]}}}  \cdot {e^{{\jmath}kN_{\rm hor}d[{\beta _{\rm hor}}(\ddot{u}  - {\mu _0}) - \dot{f}(\mu_0)(\ddot{u}-\mu_0)-\frac{1}{2}\dot{f}'(\mu_0)(\ddot{u}-\mu_0)^2]}}{\rm d}{\ddot{u}} \Bigg|\\
				= \Bigg|{e^{{\jmath}kN_{\rm hor}d[{\beta _{\rm hor}}{\mu _0} - \int_0^{{\mu _0}} {\dot{f}(\mu){\rm d}{\mu}} ]}} \cdot \int_{{\mu _0}}^{{\mu _0} + \Delta \mu_0 } {e^{{\jmath}kN_{\rm hor}d[{\beta _{\rm hor}}(\ddot{u}  - {\mu _0}) - \dot{f}(\mu_0)(\ddot{u}-\mu_0)-\frac{1}{2}\dot{f}'(\mu_0)(\ddot{u}-\mu_0)^2]}}{\rm d}{\ddot{u}}  \Bigg|\\
				= \Bigg|{e^{{\jmath}kN_{\rm hor}d[{\beta _{\rm hor}}{\mu _0} - \int_0^{{\mu _0}} {\dot{f}(\mu){\rm d}{\mu}} ]}}\Bigg| \cdot \Bigg|\int_{{0}}^{ \Delta \mu_0 } {{e^{{\jmath}kN_{\rm hor}d[{\beta _{\rm hor}}\ddot{u} - \dot{f}({\mu _0})\ddot{u}-\frac{1}{2}\dot{f}'(\mu_0)\ddot{u}^2 ]}}} {\rm d}{\ddot{u}} \Bigg|\\
				= \Bigg|\int_{{0}}^{ \Delta \mu_0 } {{e^{{\jmath}kN_{\rm hor}d[{\beta _{\rm hor}}\ddot{u} - \dot{f}({\mu _0})\ddot{u}-\frac{1}{2}\dot{f}'(\mu_0)\ddot{u}^2 ]}}} {\rm d}{\ddot{u}} \Bigg|.
			\end{array}
		\end{equation}	
		Since the array resolution increases with an increase in $kN_{\rm hor}d$, we should design $\dot{f}(\mu)$ on the condition that $kN_{\rm hor}d \to \infty$.
		If $\beta_{\rm hor}\ne \dot{f}(\mu_0)$,  the following holds as the $kN_{\rm hor}d$ tends to infinity:
		\begin{equation}\label{notequal}
			\begin{array}{l}
				\mathop {\lim }\limits_{kN_{\rm hor}d \to \infty } \Bigg|\int_{{0}}^{ \Delta \mu_0 } {{e^{{\jmath}kN_{\rm hor}d[{\beta _{\rm hor}}\ddot{u} - \dot{f}({\mu _0})\ddot{u}-\frac{1}{2}\dot{f}'(\mu_0)\ddot{u}^2 ]}}} {\rm d}{\ddot{u}} \Bigg|\\
				\overset{\text{\textcircled{1}}}{=} \mathop {\lim }\limits_{kN_{\rm hor}d \to \infty } \frac{1}{kN_{\rm hor}d} \Bigg|\int_0^{kN_{\rm hor}d\Delta {\mu _0}} {{e^{{\jmath}[\beta_{\rm hor}t-\dot{f}(\mu_0)t-\frac{1}{2}\dot{f}'(\mu_0)\frac{t^2}{kN_{\rm hor}d}]}}} {\rm d}t\Bigg|\\
				=\mathop {\lim }\limits_{kN_{\rm hor}d \to \infty } \frac{1}{kN_{\rm hor}d} \Bigg|\int_0^{kN_{\rm hor}d\Delta {\mu _0}} {{e^{{\jmath}[\beta_{\rm hor}t-\dot{f}(\mu_0)t]}}} {\rm d}t\Bigg|\\
				\le \mathop {\lim }\limits_{kN_{\rm hor}d \to \infty } \frac{2}{kN_{\rm hor}d}\\
				=0,
			\end{array}
		\end{equation}
		where \textcircled{1} is $t=kN_{\rm hor}d\ddot{u}$.
		So $\mathop {\lim }\limits_{kN_{\rm hor}d \to \infty } AFM_{\rm norm}({\beta _{\rm hor}}){|_{\ddot{u}  \in ({\mu _0},{\mu _0} + \Delta \mu_0 )}}=0$ when $\beta_{\rm hor}\ne \dot{f}(\mu_0)$.

		If $\beta_{\rm hor} = \dot{f}(\mu_0)$, then the following holds:
		\begin{equation}\label{equal}
			\begin{array}{l}
				\mathop {\lim }\limits_{kN_{\rm hor}d \to \infty } AFM_{\rm norm}({\beta _{\rm hor}}){|_{\ddot{u}  \in ({\mu _0},{\mu _0} + \Delta \mu_0 )}}\\
				=\mathop {\lim }\limits_{kN_{\rm hor}d \to \infty }\Bigg|\int_{{0}}^{ \Delta \mu_0 } {{e^{{\jmath}kN_{\rm hor}d[-\frac{1}{2}\dot{f}'(\mu_0)\ddot{u}^2 ]}}} {\rm d}{\ddot{u}} \Bigg|\\
				\overset{\text{\textcircled{2}}}{=} \mathop {\lim }\limits_{kN_{\rm hor}d \to \infty } \frac{1}{\sqrt{\dot{f}'(\mu_0)}} \Bigg|\int_{{0}}^{ \sqrt{\dot{f}'(\mu_0)}\Delta \mu_0 } {{e^{{\jmath}kN_{\rm hor}dt^2}}} {\rm d}{t} \Bigg|\\
				=\frac{K}{\sqrt{\dot{f}'(\mu_0)}}
			\end{array}
		\end{equation}
		where \textcircled{2} is $t=\sqrt{\dot{f}'(\mu_0)}\ddot{u}$, and $K$ is a varible independent to $\dot{f}(\mu)$.
		
		Since $\dot{f}(\mu)$ is an  increasing differentiable function in the range of  $[\dot{f}(0), \dot{f}(1)]$, we can find only one $\mu_0$ satisfies $\dot{f}(\mu_0)=\beta_{\rm hor}$ when $\beta_{\rm hor}\in [\dot{f}(0),\dot{f}(1)]$. Thus we can get:
		\begin{equation}\label{array factor on whole f}
			\mathop {\lim }\limits_{kN_{\rm hor}d \to \infty } AFM_{\rm norm}({\beta _{\rm hor}}) = \left\{ \begin{array}{l}
				\frac{K}{\sqrt{\dot{f}'(\mu_0)}},\;\;\;\beta_{\rm hor} \in [\dot{f}(0),\dot{f}(1)], \mu_0=\dot{f}^{-1}(\beta_{\rm hor}),\\
				\;\;\;\;0,\;\;\;\;\;\;\;\;\beta_{\rm hor} \notin [\dot{f}(0),\dot{f}(1)].
			\end{array} \right.
		\end{equation}   
		Comparing (\ref{array factor on whole f}) and (\ref{widebeamcondition}), in the wide beamforming problem we can get  $\dot{f}(0)=\psi_{{\rm hor},a}$ $\dot{f}(1)=\psi_{{\rm hor},b}$ and:
		\begin{equation}
			K\sqrt{\frac{{{\rm d}\mu }}{{{\rm d}\dot f(\mu )}}{\Bigg|_{\mu  = {\mu _0}}}} = h({\beta _{\rm hor}}),
		\end{equation} 
		where $K$ is a coefficient and $\beta_{\rm hor}=\dot{f}(\mu_0)$. We do not need to normalize the total power when selecting $h({\beta _{\rm hor}})$, as the selection of K can automatically achieve normalization. So the value of $h({\beta _{\rm hor}})$ does not actually represent the true value of $AFM_{\rm norm}({\beta _{\rm hor}})$, but rather represents the ratio relationship between the values of $AFM_{\rm norm}({\beta _{\rm hor}})$ at different $\beta _{\rm hor}$.
	Hence, the following  differential equation holds:
	\begin{equation}\label{differential equation}
		K\mu  = \int {h(\dot f(\mu )){\rm d}\dot f(\mu )}. 
	\end{equation} 
	Suppose that the solution of (\ref{differential equation}) is:
	\begin{equation}\label{solution of de}
		K\mu  = H(\dot f(\mu )) + C.
	\end{equation} 
	Putting  $\dot{f}(0)=\psi_{{\rm hor},a}$ and $\dot{f}(1)=\psi_{{\rm hor},b}$ into (\ref{solution of de}), we can get:
	\begin{equation}
		\mu  = \frac{{1 - 0}}{{H({\psi _{{\rm hor},b}}) - H({\psi _{{\rm hor},a}})}}(H(\dot f(\mu )) - H({\psi _{{\rm hor},a}})) + 0.
	\end{equation} 
	Hence, $\dot{f}(\mu)$ becomes:
	\begin{equation}\label{continues phaseshift}
		\dot f(\mu ) = {H^{ - 1}}( (H({\psi _{{\rm hor},b}}) - H({\psi _{{\rm hor},a}}))\mu  + H({\psi _{{\rm hor},a}})) .
	\end{equation}

	When $N_{\rm hor}$ tends to be infinity, $\dot{f}(\mu)$ can be considered to have  a constant value $\dot{f}(\frac{n_{\rm hor}}{N_{\rm hor}})$ in $(n_{\rm hor}-1,n_{\rm hor})$.
	Accordingly, (\ref{g&fmu}) can be simplified as:
	\begin{equation}\label{discrete phaseshift}
		g_{\rm hor}(n_{\rm hor}) = -\sum\limits_{\tau = 1}^{n_{\rm hor}} {\dot{f}(\frac{\tau}{N_{\rm hor}}} ).
	\end{equation}	
	Intuitively, for the  hierarchical codebook, the best wide beam power distribution is the  power evenly distributed in the beam range. Hence, $h(\beta_{\rm hor})$ is a constant function, and  $\dot{f}(\mu)$ can be simplified as:
	\begin{equation}\label{fx_ofwidebeam}
		\dot{f}(\mu)=(\psi_{{\rm hor},b}-\psi_{{\rm hor},a}) \mu + \psi_{{\rm hor},a}.
	\end{equation}
	Since (\ref{fx_ofwidebeam}) is a linear function,  we call this algorithm as the  none-constant phase shift difference (NCPD) algorithm.
	Therefore $g_{\rm hor}(n_{\rm hor})$ can be expressed  as:
	\begin{equation}\label{anasou}
		g_{\rm hor}(n_{\rm hor}) = -(\sum\limits_{\tau = 1}^{n_{\rm hor}}( {({\psi _{{\rm hor},b}} - } {\psi _{{\rm hor},a}})\frac{\tau}{N_{\rm hor}} + {\psi _{{\rm hor},a}}))\\
		=-(\frac{{{n_{\rm hor}}(n_{\rm hor} + 1)}}{{2N_{\rm hor}}}({\psi _{{\rm hor},b}} - {\psi _{{\rm hor},a}}) + n_{\rm hor}{\psi _{{\rm hor},a}}).
	\end{equation}
	\newtheorem{remark}{Remark}
	\begin{remark}
		If $\dot{f}(\mu)$ is not monotonic, then it can be viewed as a piecewise function, where $\dot{f}(\mu)$ is monotonic in each segment, and the conclusions described in this paper can still be applied. Therefore, in order to simplify the expression, this paper discusses only the case where $\dot{f}(\mu)$ is a monotonic function.
	\end{remark}
	\begin{remark}
		The above algorithm is developed based on the constant modulus constraint. So, the conclusions work not only on IRS but also on analog beamforming of active equipments.
	\end{remark}

	\section{Design of Hierarchical Codebook on IRS}
	Since the  cascaded channel direction is determined by using  $\beta_{\rm hor}$ and $\beta_{\rm ver}$, and the hierarchical codebook on them are the  same, in this section we first introduce the hierarchical codebook in a single direction, and then propose two  beam training algorithms.
	
	\subsection{Hierarchical Codebook on One Direction}
\begin{figure}[!ht]
		\centering
		\includegraphics[scale=1]{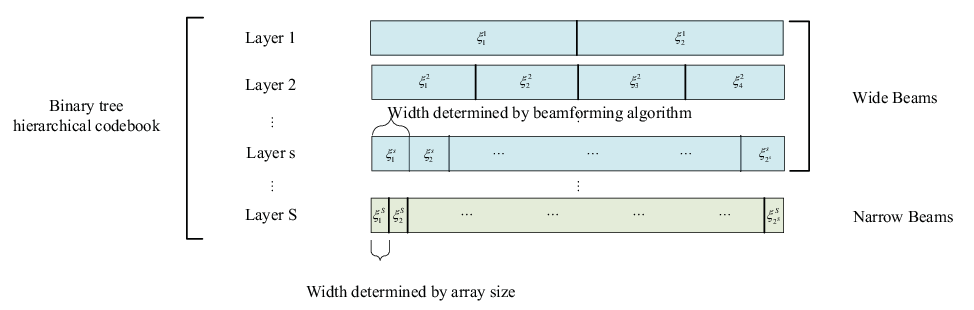}
		\caption{ Hierarchical codebook structure.}     \label{codebook_u}
	\end{figure}

	Suppose that the  hierarchical codebook has $S$  layers, and  $s$ is its layer index, where  $s=1,2,  \cdots  S$. We use ${\bm\Xi}_s$ to denote the codebook of layer $s$. ${\bm \Xi}_S$ is the bottom layer with narrow beams, other layers use wide beams. $\bm \Xi_1$ has two beams and ${\bm \Xi}_2$ has four beams, and ${\bm \Xi}_s$ has $2^s$ beams. From \cite{DBLP:journals/tvt/NingCCDF21}, we know that there are $2N_{\rm hor}$ narrow beams in the  ${\bm h}_{\rm hor}$ direction, and $2^S=2N_{\rm hor}$. We can find that the hierarchical codebook needs $\lceil log_2 N_{\rm hor}\rceil$ layers. The $s$th layer of the hierarchical codebook can be expressed as:
	\begin{equation}
		{\bm \Xi}_s=[{\bm\xi}_1^s,{\bm\xi}_2^s,\cdots {\bm \xi}_{2^s}^s],
	\end{equation}
	where  ${\bm \xi}_i^s$ is  a $N_{\rm hor}$-dimensional column vector.
	For the bottom layer,  the beams can be expressed as:
	\begin{equation}
		{\bm \xi}_i^S[n_{\rm hor}]= e^{{\jmath}kd(n_{\rm hor}-1)\frac{2(i-N_{\rm hor})-1}{N_{\rm hor}}}.
	\end{equation}
	For any other layer $s$ $(s \ne S)$, we first  need to confirm the range of the directions of the beams. Suppose that  $(\psi_{a,{\bm \xi}_i^s},\psi_{b,{\bm \xi}_i^s})$ is the directional range of beam ${\bm \xi}_i^s$
	\begin{equation}\label{xiab}
		\begin{split}
			\psi_{a,{\bm \xi}_i^s}=-2+\frac{4}{2^s}(i-1),\;
			\psi_{b,{\bm \xi}_i^s}=-2+\frac{4}{2^s}i.
		\end{split}
	\end{equation}
	Putting them into (\ref{anasou}), we can get the beam ${\bm \xi}_i^s$ $(s \ne S)$ as:
	\begin{equation}\label{xi}
		\begin{split}
			&{\bm \xi}_i^s[n_{\rm hor}]={\rm exp}\{-{\jmath}kd[\frac{n_{\rm hor}(n_{\rm hor}+1)}{2N_{\rm hor}} \cdot \frac{4}{2^s}-2+ \frac{4}{2^s}(i-1)]\}.\\
			&s=1,2, \cdots S-1,\;\;
			i=1,2, \cdots 2^s.
		\end{split}
	\end{equation}
	For the ${\bm h}_{\rm ver}$ direction, the hierarchical codebook are exactly the same as that in  the ${\bm h}_{\rm hor}$ direction. Fig. \ref{codebook_u} shows  the structure  of the codebook.

	\subsection{Beam Training Scheme for Jointment Searching}
	\begin{figure}
		\centering
		\includegraphics[scale=1]{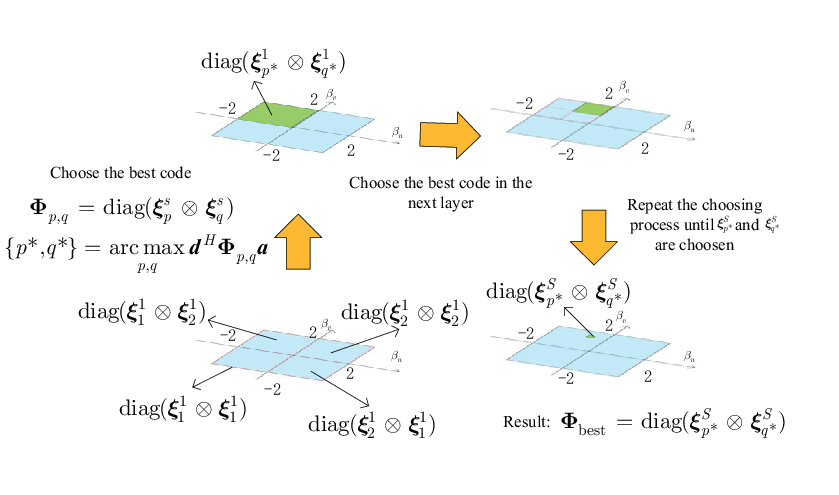}
		\caption{ JS beam training process.}     \label{JS}
	\end{figure}
	In this sub-section, we will propose a searching scheme to search ${\bm h}_{\rm hor}$ direction and ${\bm h}_{\rm ver}$ direction at the same time which is called jointment searching (JS) scheme. The phase shift matrix  of the IRS is searched by using the above-described hierarchical codebook.  According to (\ref{array factor final}), the beam direction of the IRS can be divided into the ${\bm h}_{\rm hor}$ direction and the ${\bm h}_{\rm ver}$ direction, and the JS search scheme will search in these two directions synchronously. In the $s$th layer search, codewords in ${\bm \Xi}^1$ are used in the ${\bm h}_{\rm hor}$ and ${\bm h}_{\rm ver}$ directions. According to (\ref{guv}),  the phase shift matrix of IRS in the  $s$th layer search can be written as:
	\begin{equation}\label{JSPhi}
		{\bm \Phi}_{p,q}={\rm diag}({\bm \xi}^s_p \otimes {\bm \xi}^s_q).
	\end{equation}
	Then, according to (\ref{received signal}), the received signal is obtained, and the received signal with the largest power is selected, that is, the indices  of the  best code of $s$th layer  is selected. According to (\ref{dphia}) we know:
	\begin{equation}\label{JSbest}
		\{p^*,q^*\}=\mathop{\arg\max}\limits_{p,q}{\bm d}^H {\bm \Phi}_{p,q}{\bm a} .
	\end{equation}
	At this point, the searching procedure of the $s$th layer is finished, and then we can search the codes in the next layer. Finally we can get the best phase shift matrix of IRS after the bottom  layer of the hierarchical codebook are searched. The details of the proposed algorithm are summarized in Algorithm  \ref{alg1}.

	\makeatletter
	\newcommand{\removelatexerror}{\let\@latex@error\@gobble}
	\makeatother
	%\begin{figure}[!t]\label{alg:1}
	\renewcommand{\algorithmicrequire}{\textbf{Input:}}
	\renewcommand{\algorithmicensure}{\textbf{Output:}}
	\removelatexerror
	\begin{algorithm}
		
		\caption{JS codebook design}
		\begin{algorithmic}[1]\label{alg1}
			\STATE {Initialize  the  layer index $s=1$ and $p^*=1$.}
			\REPEAT{}
			\STATE{Compute ${\bm \Phi}_{p,q}$ by (\ref{JSPhi}), where $p \in \{2p^*-1,2p^*\}$, $q \in \{2q^*-1,2q^*\}$.}
			\STATE{Update $p^*$ by choosing the indices of the best codes of the $s$th layer by (\ref{JSbest}).}
			\STATE{Update $s=s+1$.}
			\UNTIL{$s>S$.}
			\STATE {${\bm \Phi}_{\rm best}={\rm diag}({\bm \xi}^S_{p^*}\otimes{\bm \xi}^S_{q^*})$.}

		\end{algorithmic}
	\end{algorithm}
	\vspace{-0in}%	

	\subsection{Direction Wise Searching Scheme}
	In this section, we  propose another beam searching scheme on IRS named direction wise searching  (DWS) scheme. The difference between the DWS and the JS scheme is that the DWS scheme first searches the ${\bm h}_{\rm hor}$ direction, and then starts searching in the ${\bm h}_{\rm ver}$ direction after determining the codeword in the ${\bm h}_{\rm hor}$ direction. Therefore, the whole search process is divided into two stages. In the first stage, the $\bm \Xi$ codebook is used for searching in the ${\bm h}_{\rm hor}$ direction, and an omnidirectional codeword ${\bm \xi}_{\rm ver}$ is used in the ${\bm h}_{\rm ver}$ direction. According to formula (\ref{anasou}), we can get:
	\begin{equation}
		{\bm \xi}_{\rm ver}[n_{\rm ver}]=-\frac {2n_{\rm ver}(n_{\rm ver}+1)}{N_{\rm ver}}+2n_{\rm ver}, n_{\rm ver}=1,\cdots,N_{\rm ver}.
	\end{equation}
	In the $s$th layer search, codewords of  ${\bm \Xi}^s$ are used in the ${\bm h}_{\rm hor}$ direction, and ${\bm \xi}_{\rm ver}$ is used in the ${\bm h}_{\rm ver}$ direction. Thus the codes of $s$ layer search in stage 1 can be written as:
	\begin{equation}\label{DWSPhi}
		{\bm  \Phi}_{p,0}={\rm diag}({\bm \xi}^s_p \otimes {\bm \xi}_{\rm ver}).
	\end{equation}	
	Then we select the index of the best codewords by the following:
	\begin{equation}\label{DWSp*}	
		p^*=\mathop{\arg\max}\limits_{p}{\bm d}^H  {\bm \Phi}_{p,0} {\bm a} .
	\end{equation}	
	Then repeat the above searching procedure until the bottom layer of $\bm \Xi$. At this time, the searching procedure  in the ${\bm h}_{\rm hor}$ direction is completed, and the second stage begins to search  for the ${\bm h}_{\rm ver}$ direction.

	\begin{figure}[!ht]\label{DWS}
		\centering
		\includegraphics[scale=1]{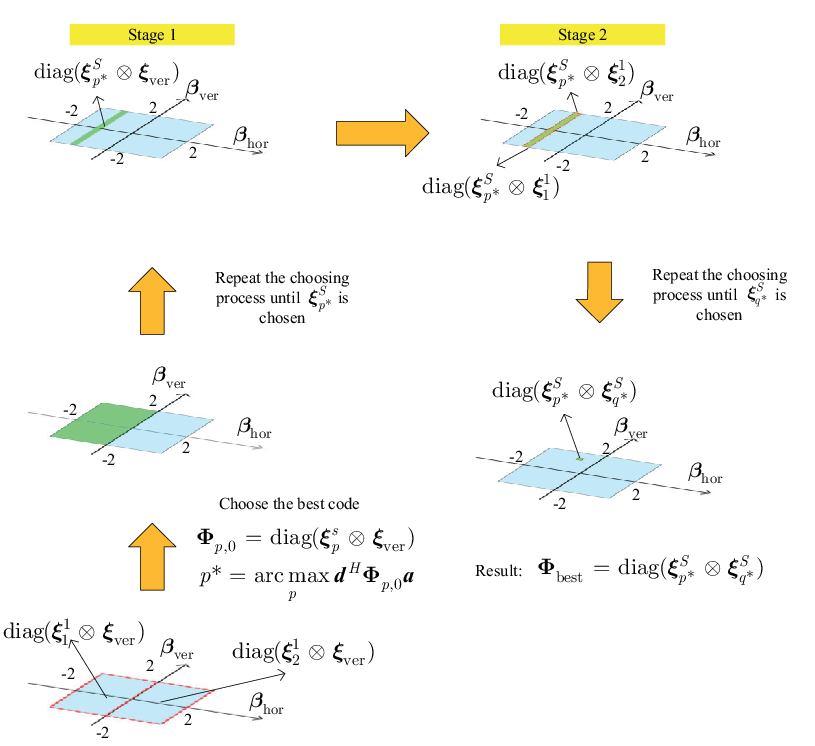}
		\caption{ DWS beam training process.}    
	\end{figure}

	%	\makeatletter
	%	\newcommand{\removelatexerror}{\let\@latex@error\@gobble}
	%	\makeatother
	%	%\begin{figure}[!t]\label{alg:1}
	%	\renewcommand{\algorithmicrequire}{\textbf{Input:}}
	%	\renewcommand{\algorithmicensure}{\textbf{Output:}}
	
	\begin{algorithm}
		
		\caption{DWS codebook design}
		\begin{algorithmic}[1]\label{alg2}
			\STATE {Initialize the layer index $s=1$ and $p^*=1$ for stage 1.}
			\REPEAT
			\STATE{Compute ${\bm \Phi}_{p,0}$, where $p \in \{2p^*-1,2p^*\}$ by (\ref{DWSPhi}).}
			\STATE{Update $p^*$ by choosing the index of the best codes of $s$th layer by (\ref{DWSp*}).}
			\STATE{Update the layer index: $s=s+1$.}
			\UNTIL{$s>S$.}
			\STATE{Initialize the layer index $s=1$ and $q^*=1$ for stage 2.}
			\REPEAT
			\STATE{Compute ${\bm \Phi}_{p^*,q}$, where $q \in \{2q^*-1,2q^*\}$ by (\ref{DWSPhi2}).}
			\STATE{Update $q^*$ by choosing the index of the best codes of $s$th layer by (\ref{DWSq*}).}
			\STATE{Update the layer index: $s=s+1$.}
			\UNTIL{$s>S$.}
			\STATE {${\bm \Phi}_{\rm best}={\rm diag}({\bm \xi}^S_{p^*}\otimes {\bm \xi}^S_{q*})$.}

		\end{algorithmic}
	\end{algorithm}
	\vspace{-0in}%

	In the stage 2, the ${\bm h}_{\rm hor}$ direction uses the selected codewords in the stage 1, that is ${\bm \xi}^S_{p^*}$, and the ${\bm h}_{\rm ver}$ direction uses $\bm \Xi$ for hierarchical search. For the $s$th layer search, the phase shift matrices of IRS can be written as:
	\begin{equation}\label{DWSPhi2}	
		{\bm \Phi}_{p^*,q}={\rm diag}({\bm \xi}^S_{p^*} \otimes {\bm \xi}^s_q).
	\end{equation}	
	In the same way, the optimal $q$ is selected according to the received signal:
	\begin{equation}\label{DWSq*}	
		q^*=\mathop{\arg\max}\limits_{q}{\bm d}^H  {\bm \Phi}_{p^*,q} {\bm a} .
	\end{equation}	
	Repeat the above searching procedure until the codes in the  last layer of the hierarchical codebook are searched. Finally we get the best codeword ${\bm \Phi}_{\rm best}={\rm diag}({\bm \xi}^S_{p^*} \otimes {\bm \xi}^S_{q^*})$. The details of the proposed algorithm are summarized in Algorithm  \ref{alg2}.
	
	\subsection{Performance Comparison of Algorithm and Analysis of Time Complexity}	
	In this section, the JS and DWS schemes are compared in terms of their time complexity and designing  flexibility. 
	
	When the IRS array is a square matrix, i.e., $N_{\rm ver}=N_{\rm hor}$, and both JS and DWS schemes use binary tree structured codebook for searching, their time complexities become the same, which is $\mathcal{O}(\lceil{\rm log}_2N\rceil)$. However, if the hierarchical codebook are changed from a binary tree structure to an $M$-ary tree structure, the time complexity of the JS scheme becomes $\mathcal{O}(M^2\lceil{\rm log}_M N\rceil)$, and that of the DWS scheme becomes  $\mathcal{O}(2M\lceil{\rm log}_M N\rceil)$. Accordingly, the DWS scheme takes less time.
	
	In terms of the  design freedom and algorithmic flexibility, the DWS algorithm  is better than the JS algorithm. In the JS scheme, although the codebooks used in the ${\bm h}_{\rm hor}$ and ${\bm h}_{\rm ver}$ directions are binary tree structures, the entire JS algorithm actually searches in a quad-tree structure as the ${\bm h}_{\rm hor}$  and  ${\bm h}_{\rm ver}$ directions are searched at the same time. Hence, we can conclude that the JS scheme cannot implement a search structure with the branch  number less than 4 or an odd number of branches.  The DWS scheme can realize the search structure of any number of branches. In addition, the JS scheme can  be used only when the IRS array is a square array, while the DWS scheme does not have this limitation. Therefore, the DWS scheme is more flexible.
	
	However, in some special scenarios, it is not necessary  to complete all the searching processes, i.e., it is not necessary to search for any narrow beam at the bottom layer of the hierarchical codebook. For example, when the array size is practically large, the power required for communication can already be achieved by using  wide beams in the middle layer of the hierarchical codebook. In this situation, completing the entire searching process will waste time and increases delay. In addition, if  UEs moves slowly, the use of a wider beam can avoid the problem of frequent beam switching. In the above scenarios, we need to terminate the searching progress before the beam training process is completed, and then directly use the beam of the intermediate result of the beam training process  for communication. The searching process of the JS scheme needs to simultaneously narrow the scopes of the ${\bm h}_{\rm hor}$ and ${\bm h}_{\rm ver}$ directions, while that of the DWS scheme needs to first narrow the scope of the ${\bm h}_{\rm hor}$ direction, and then that of the ${\bm h}_{\rm ver}$ direction. Therefore, the intermediate results of the DWS search are not suitable for the above scenarios, while the  JS scheme can be interrupted at any time.
	
	It is worth mentioning that the JS and DWS schemes are not "either-or" choices. Sometimes we can use the JS and DWS schemes simultaneously for beam training to combine the individual advantages of the two schemes.  For example, we can perform the JS search first, and stop when the JS search is not completed, and then switch to the DWS scheme for continuing the beam training in the area to be searched. This not only takes advantage of the feature that that can interrupt the JS scheme  in the middle, but also takes advantage of the lower time complexity of the DWS scheme.
	
	\section{Performance and Time Complexity Analysis }
		In this section, we will explain the difference between NCPD algorithm and traditional beamforming algorithm from the perspective of design principle, performance and time complexity, and introduce the main advantages of NCPD algorithm on large arrays.
		
		The traditional beamforming design algorithms are designed directly based on the discrete phase shift matrix, while the NCPD algorithm designs the continuous phase shift function first, and then obtains the final discrete beamforming solution through sampling progress.  The time spent in obtaining discrete beamforming solutions from continuous phase shift function sampling can often be ignored. However, in order to show the performance of NCPD more comprehensively, we have written down the time complexity of both forms of solutions in Table \ref{t1}. Since (\ref{continues phaseshift}) is irrelevant to $N$, the time complexity of continuous phase shift function is $\mathcal{O}(1)$, while the discrete  result in (\ref{discrete phaseshift}) is $\mathcal{O}(N^2)$. 
		\begin{table}
			\centering
			\caption{\textsc{Time complexity comparison.}}
			\begin{tabular}{|c|c|} 
				\hline 
				Algorithm  & Time complexity  \\
				\hline
				Iterative optimization method \cite{DBLP:journals/wcl/MaZZ22} & $\mathcal{O}(N^{6.5})$  \\
				\hline
				Riemannian optimization \cite{DBLP:journals/wcl/FanZH19}& $\mathcal{O}(N^4)$ \\
				\hline
				Proposed NCPD (continuous phase shift function) & $\mathcal{O}(1)$ \\
				\hline
				Proposed NCPD (final result) & $\mathcal{O}(N^2)$ \\
				\hline
			\end{tabular}
			\label{t1}
		\end{table}
		
		In fact, the optimized beamforming design method and the NCPD algorithm are two completely different design logics. From Table \ref{t2}, we can see that on large arrays, optimization algorithms and NCPD algorithms have high flexibility and performance, while on-off elements algorithm, beam combination algorithm, and NCPD algorithm have lower computational time. In general, NCPD algorithm  has obvious advantages on large arrays. Among the above algorithms, only NCPD algorithm  has both high beam design freedom and low time complexity. Moreover, the larger the array, the better is the performance of the NCPD algorithm, and the more obvious is the advantages.

	\begin{table*}[!httbp]
		\centering
		\caption{\textsc{Comparison of structure between optimization algorithm and NCPD algorithm.}}
		\begin{tabularx}{\textwidth}{|p{2.5cm}|p{6.7cm}|p{6cm}|}
			\toprule
			\textbf{ Algorithm } &  \textbf{Performance } & \textbf{Time }  \\
			\hline
			Optimization methods \cite{DBLP:journals/wcl/FanZH19,DBLP:journals/wcl/MaZZ22} & Depends on the construction of the objective function and constraint function, as well as the setting of iteration step size and number of iterations. & The larger the array, the longer is  the operation time.\\
			\hline
			Sub-array and beam combination methods \cite{DBLP:journals/wcl/YouZZ20,DBLP:journals/icl/SinghK22,DBLP:journals/twc/XiaoHXX16}  & There are three factors that determine performance: the first is the number of sections to be divided, the second is the array size, and the third is the width and shape of the target beam. & The difference in computation time between large and small arrays is not significant.  \\
			\hline
			On-off elements & RF chains can be used on active arrays to compensate for the losses caused by closing elements, while passive arrays cannot compensate for this part of the losses, resulting in poor performance. Meanwhile, this algorithm has low design flexibility. & Operation time is independent of array size.\\
			\hline
			Proposed NCPD & Performance is only determined by the size of the array, and the larger the array, the better is the performance. & The  time of a continuous form solution is not related to the size of the array, but only to the complexity of the target beam. The discrete form solution only increases the sampling time above it, and the sampling time can often be ignored. \\
			\bottomrule
		\end{tabularx}%
		\label{t2}%
		
	\end{table*}%

	\section{Simulation Results }
	In this section, the theoretical results described above are verified, and the performance of the NCPD algorithm is show in forming  beams of different widths on IRS of different sizes. Since the ${\bm h}_{\rm hor}$  and ${\bm h}_{\rm ver}$ directions are independent of each other and the beamforming designs in these two directions are completely equivalent, we  show the array factor in one direction only. Therefore,  the number of elements refers to the value of either $N_{ver}$ or $N_{hor}$. Since the value   of $\beta_{\rm hor}$ is $(-2,2)$, we use $\frac{\pi}{2}\beta_{\rm hor}$ as the independent variable for forming the array factor in the polar coordinate system. But the concept of "beam width" still refers to the width of range of $\beta_{\rm hor}$. For example if the range of  $\beta_{\rm hor}$ in a wide beam is $(0,0.5)$, then the width of that wide beam will be $0.5$. 
	
	In this section, we  use the beam combination method as a benchmark algorithm for wide beamforming problems. In the beam combination method, IRS is divided into many sub-arrays, each of which forms a narrow beam, and then these narrow beams are superimposed to form a wide beam. 
	%Many articles give some reasonable ways of dividing sub-arrays, but these methods have one thing in common, that is, the way of dividing the sub-arrays should be artificially changed according to the size of the array and the width of the target beam to be formed. This is because if the sub-array division method is fixed, with the increase of the size of the array and the increase of the width of the formed target beam, the deep depression of the beam will become more and more serious. 
	In this experiment, we choose two beam combination methods, in which four sub-beams are used to synthesize wide beams of 4-combination and sixteen sub-beams are used to synthesize wide beams of 16-combination. 
	
	\subsection{Comparison under Different Array Sizes.}
\begin{figure}
		\centering
		\subfigure[  4-combination algorithm.] {\includegraphics[width=.6\textwidth]{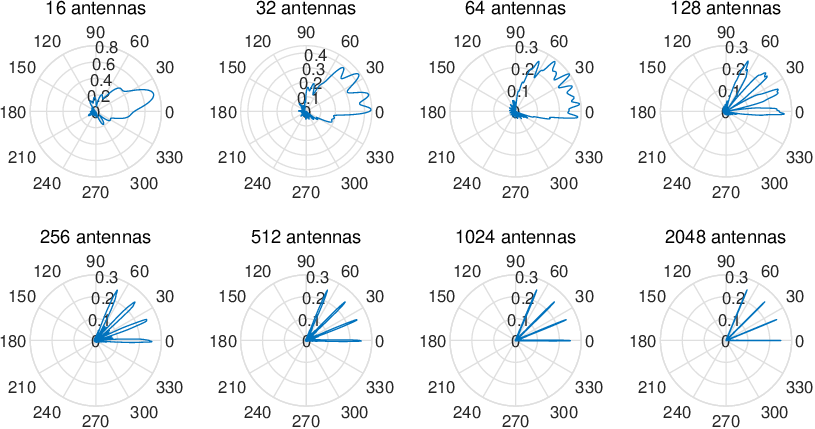}}\\
		\subfigure[  16-combination algorithm.] {\includegraphics[width=.6\textwidth]{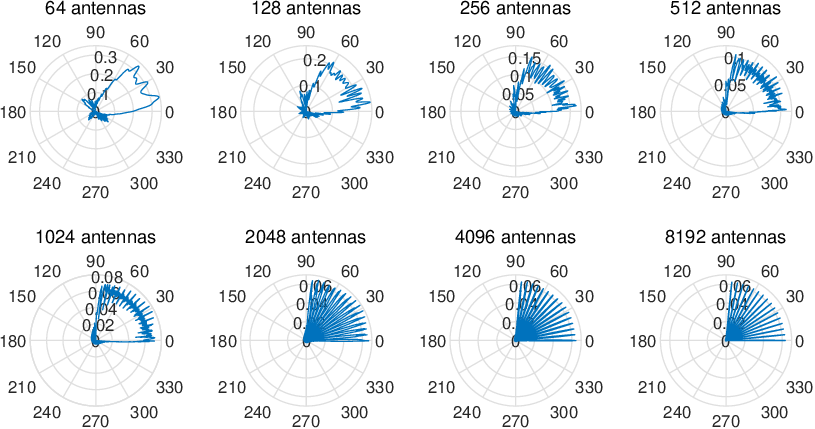}}\\
		\subfigure[  NCPD algorithm.] {\includegraphics[width=.6\textwidth]{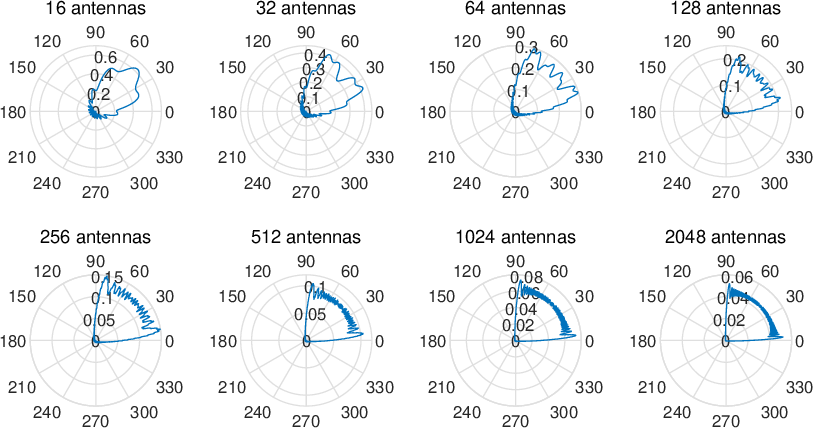}}	
		\caption{  Comparison of beam combination algorithm and NCPD algorithm in different sizes of IRS. }   \label{f1}
	\end{figure}

	Fig. \ref{f1} shows the plots of the results of the  beam combination and NCPD algorithms obtained in forming beams of width $1$ at different array sizes. Figs. \ref{f1}(a) and \ref{f1}(b) show those of  the 4-combination and 16-combination algorithms and  NCPD algorithm, respectively. As can be seen from Fig. \ref{f1}(a), the performance of the 4-combination algorithm is the best when the total number of elements is 64. Due to the insufficient elements in the sub-arrays when the total number of elements is less than 64, the formed sub-beam is not accurate enough, which means that the main lobe of the sub-beam is relatively wide and the side lobes are relatively high, thus enabling the main lobes of the four sub-beams to get connected with each other. However, when the total number of elements is greater than 64, it can be seen that each sub-beam is more accurate and its main lobe  is narrower, resulting in a deep depression among several sub-beams. 
	
	From Fig. \ref{f1}(b), we can see that when the total number of elements is 64, the performance of the 4-combination algorithm is better than that of the 16-combination algorithm. It means that when the total number of elements is small, we need to use fewer sub-beams to achieve a beam combination. The main reason of this phenomenon is that when the total number of elements is small, the number of elements allocated to each sub-beam in the 4-combination algorithm is more than that in the 16-combination algorithm, which makes each sub-beam in the combination algorithm  more accurate and to have  smaller side lobes. When the total number of elements increases to 128, we can see that the 16-combination algorithm does not have the deep depression phenomenon that the 4-combination algorithm does. When the total number of elements increases to 1024, the performance of the 16-combination algorithm becomes the best, which means that the more the total number of elements, the more sub-arrays are needed to form more sub-beams for combining  wide beams. However, when the total number of elements is greater than 1024, the 16-combination algorithm also has a deep depression phenomenon. Since the more the number of elements, the more the power of the sub-beams is concentrated in their center direction,  no matter how many sub-beams are used for combination, but the phenomenon of deep depression cannot be avoided.
	
	From Fig. \ref{f1}(c), we can see that the NCPD algorithm is slightly better than the beam combination algorithm when the number of elements is small. However, when the number of elements increases, the NCPD algorithm does not achieve deep depression, and its performance is much better than that of the beam combination algorithm. With the increase in the number of elements, the side lobes of the beam formed by the NCPD algorithm become smaller, the main lobe becomes flatter and the edge of the main lobe becomes sharper, which gradually tends to our design goals. This shows that the NCPD algorithm can solve the problem of deep recesses when the number of elements increases. The NCPD algorithm is applicable to arrays of any size, and there is no need to artificially change the beamforming algorithm for the array size. Therefore, we can conclude that the NCPD algorithm is robust to the number of elements. Therefore, the time complexity of the NCPD algorithm is $\mathcal{O}(1)$, and the performance and speed in solving the IRS phase shift matrix in different scenarios are significantly better than those of the beam combination algorithm.
	
	\subsection{Comparison under Different Beam Width.}
\begin{figure}
		\centering
		\subfigure[4-combination algorithm.] {\includegraphics[width=.7\textwidth]{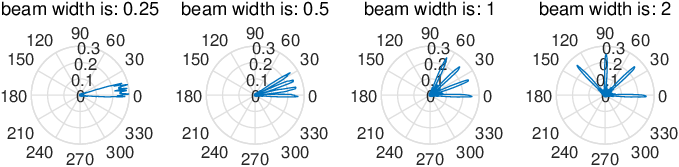}}\\
		\subfigure[16-combination algorithm.] {\includegraphics[width=.7\textwidth]{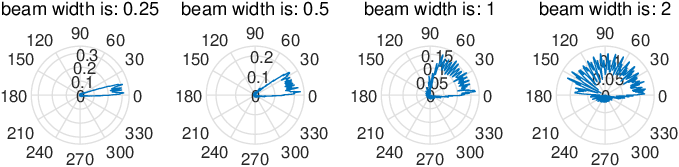}}\\
		\subfigure[NCPD algorithm.] {\includegraphics[width=.7\textwidth]{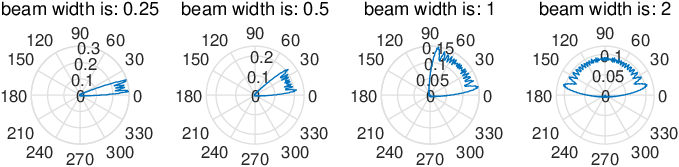}}	
		\caption{  Comparison of beam combination algorithm and NCPD algorithm when forming different width beams. }  \label{f2} 
	\end{figure}

	Fig. \ref{f2} shows the performance comparison when the beam combination  and NCPD algorithms form wide beams with widths of 0.25, 0.5, 1, and 2 on 256 elements, respectively. Figs. \ref{f2}(a)-\ref{f2}(c) show the performance of the 4-combination, the 16-combination, and the NCPD algorithms, respectively. We can see that both  4-combination and 16-combination algorithms have deep depressions when they synthesize  beams with large widths, and the deep depression of the 4-combination algorithm is more serious. The performance of the NCPD algorithm is slightly better than those of  the 4-combination and 16-combination algorithms when they form a narrow beam. However, when a wide beam is formed, there is no deep depression, and the effect is much better than those of  the two beam combination algorithms. From this, it can be concluded that the NCPD algorithm is also robust when forming beams of different widths.

	\subsection{Performance of NCPD Algorithm on Giant IRS}	
\begin{figure}
		\centering
		\includegraphics[scale=0.5]{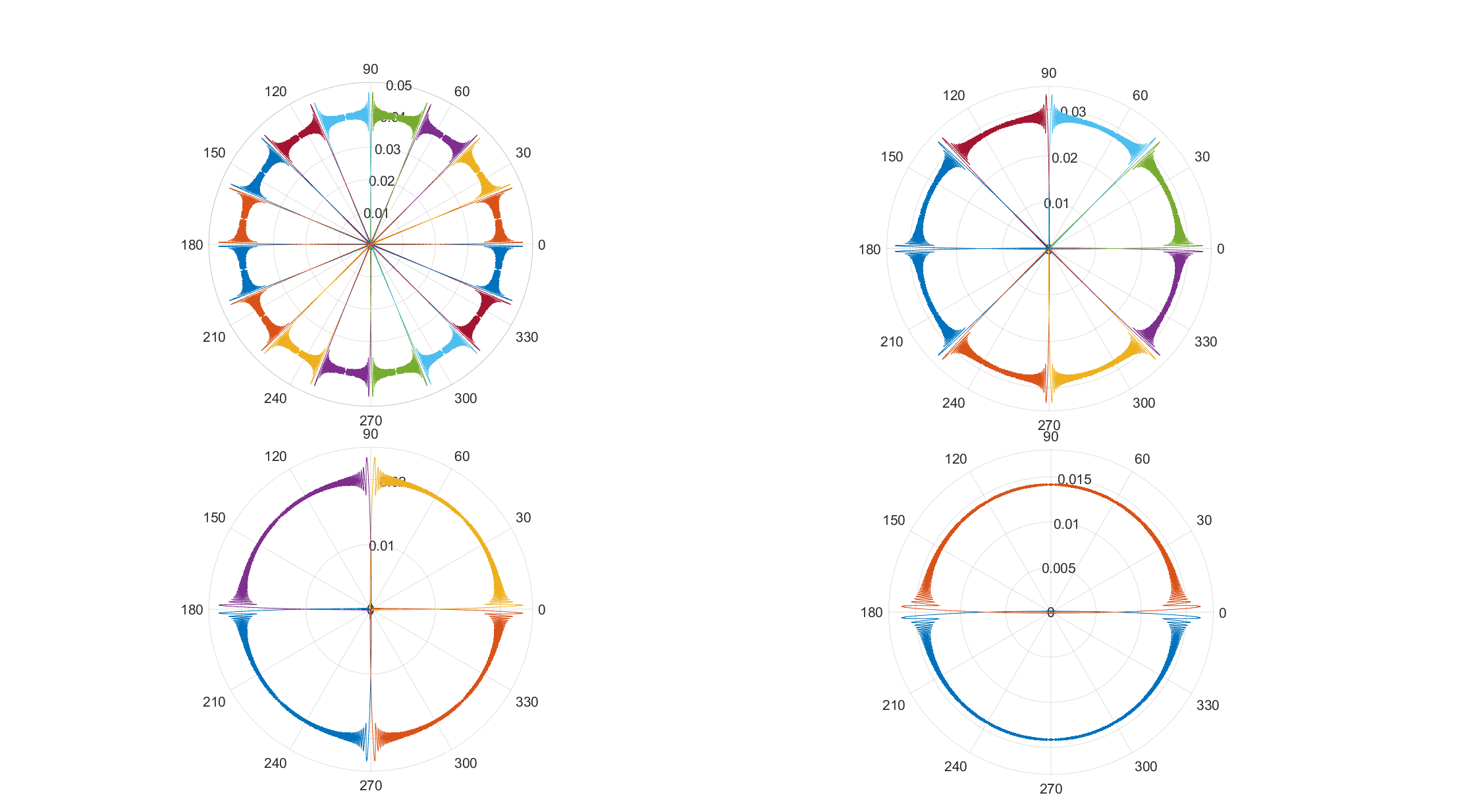}
		\caption{  The beams formed by NCPD algorithm on 10000 element IRS. } \label{f3}  
	\end{figure}

	Fig. \ref{f3} shows the results of the NCPD algorithm forming beams of different widths for the number of elements of 10000, where we can see that the beam has almost no side lobes, its edge is sharply cut off, and the power in it is flatly distributed. This shows the superiority of the NCPD algorithm in dealing with beamforming problems on large-scale arrays. We found that at the edge of each beam, there is an overshoot phenomenon, which is the Gibbs phenomenon in beamforming. The Gibbs phenomenon in beamforming arises as IRS consists of discrete elements rather than a continuous reflector. Therefore, the beam-forming error caused by the Gibbs phenomenon is an inherent error of this problem, which cannot be eliminated by changing the algorithm.

	\begin{figure}
		\centering
		\includegraphics[scale=0.45]{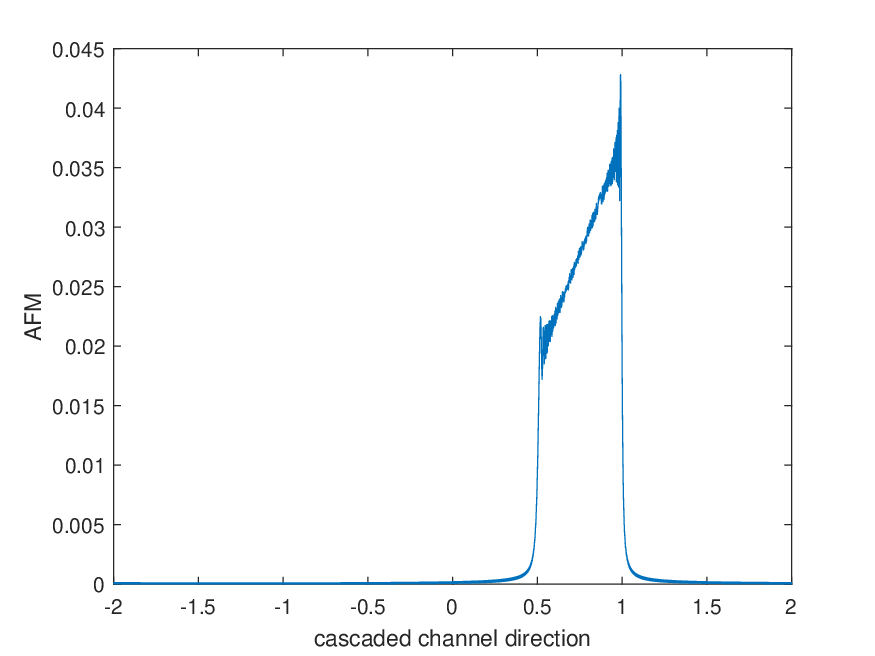}%beamshape.eps
		\caption{ The trapezoidal beams formed by NCPD algorithm on 10000 element IRS.} \label{beamshape}  
	\end{figure}

		Figure \ref{beamshape} shows the NCPD algorithm forming a trapezoidal beam on IRS with 10000 elements. The control function for beam shape is: $h(\beta)=\beta, \beta \in (0.5,1)$. Trapezoidal beams cannot be used for hierarchical codebooks. This simulation is only to demonstrate the ability of the NCPD algorithm in controlling beam shapes. It can be seen that this trapezoidal beam still has the characteristic of low sidelobes, and its main lobe shape is completely consistent with the pre-set shape.

	\subsection{Comparison of Misalignment Rates between NCPD  and Beam Combination Algorithms}
	
	Next, we  compare the misalignment rates of wide beams formed by the NCPD, beam combination algorithms and the beam in the ideal case. The beam combination algorithm is the BMW-SS algorithm proposed in \cite{DBLP:journals/twc/XiaoHXX16}. In the BMW-SS algorithm, a binary tree is also used for implementing a hierarchical codebook, in which the bottom layer also uses a narrow beam. In the case of wide beams, the BMW-SS algorithm  uses $2^{\lfloor \frac{l+1}{2}\rfloor}$ sub-beams for synthesis, where $l$ represents the number of layers away from the bottom layer. For example, when forming a wide beam of the penultimate layer, $l=1$, that is, the array is divided into two sub-arrays, and two sub-beams are respectively realized, and then the two sub-beams are synthesized into the required wide beam.  Since the wider the beam, the smaller is the average power in each direction,  the codeword of the top layer has the highest probability of misalignment so under the same signal-to-noise ratio (SNR). Therefore, the misalignment  rate of the hierarchical  codebook mainly depends on that of the first layer of codewords. Hence, we  show the comparison of the misalignment only of the first layer of codewords. The ideal case means the ideal beams in the coverage of $(-2,0)$ and $(0,2)$ with the power of $2N_{ver}$ or $2N_{hor}$, which are the target wide beams of the first layer of the hierarchical codebook. It is worth mentioning that there're no codewords corresponding to ideal case. We just suppose the formed beams are in the ideal shape and test the misalignment rate with the use of beams in this shape.
	
	We conducted two experiments to show the comparison between the NCPD  and the BMW-SS algorithms under different IRS received SNRs and different IRS array sizes. The IRS received SNR refers to the ratio of the total received power of  IRS to the noise power of the cascaded channel. Since we choose the total received SNR of  IRS rather than the received SNR on a single element, we can exclude the gain caused by the increase in the IRS array size, and  show only the beamforming gain of  IRS.
	
	\begin{figure} 
		\centering
		\includegraphics[width=.5\textwidth]{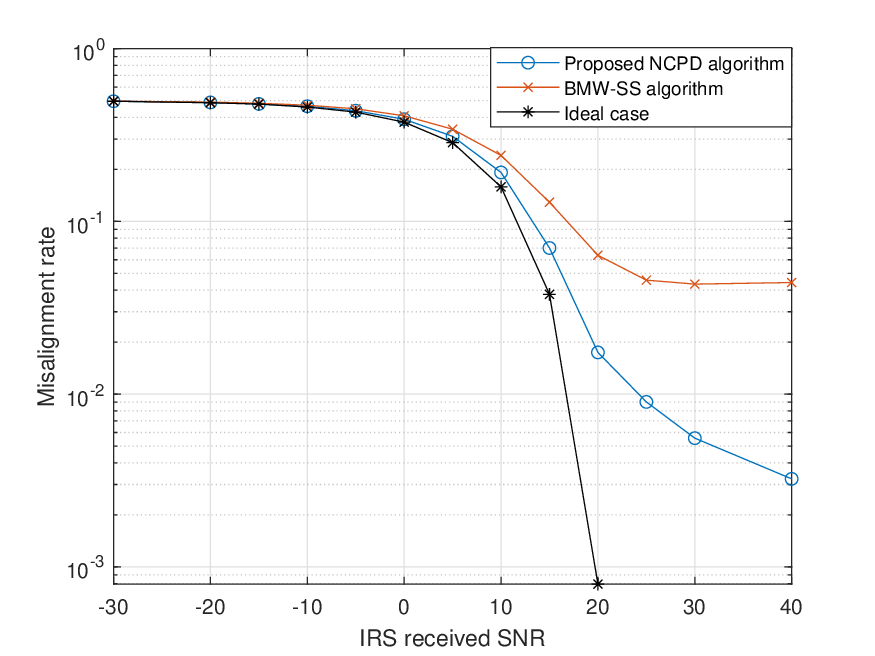}
		\caption{  Comparison of beam misalignment rates under different IRS received SNR. }  \label{f4} 
	\end{figure}
	
	Fig. \ref{f4} shows the misalignment curves of the first-layer codewords of the NCPD  and the BMW-SS algorithms and the ideal case under different IRS received SNRs on the uniform linear array (ULA) of 256 elements. We can see that when  SNR is less than 0dB, the misalignment rates of the two algorithms are high. This is because the noise power at this stage is too large, and  IRS cannot use the beamforming algorithm to offset the noise interference. But when  SNR is greater than 0dB, the misalignment rate of the NCPD algorithm is significantly lower than that of the BMW-SS algorithm  as the power distribution in the beam formed by the NCPD algorithm is more uniform. At the same time, we noticed that when  SNR is greater than 30dB, the misalignment rate of the NCPD algorithm is still declining, while that in the BMW-SS algorithm is almost unchanged. This is because when  SNR is high, the deep depression problem becomes the main reason for limiting the performance of the BMW-SS algorithm, while the performance of the NCPD algorithm is gradually improved with increasing SNR as it has no deep depression problem.
	
	\begin{figure}
		\centering
		\includegraphics[width=.5\textwidth]{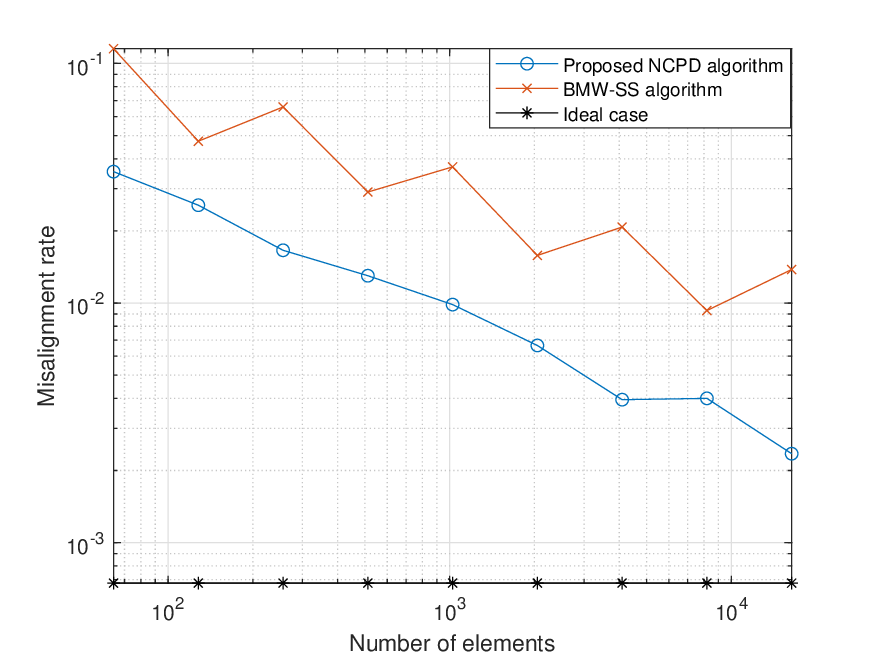}
		\caption{  The misalignment rates on different IRS size when IRS received SNR is 20 dB. }  \label{f5} 
	\end{figure}
	
	\begin{figure}
		\centering
		\includegraphics[width=.5\textwidth]{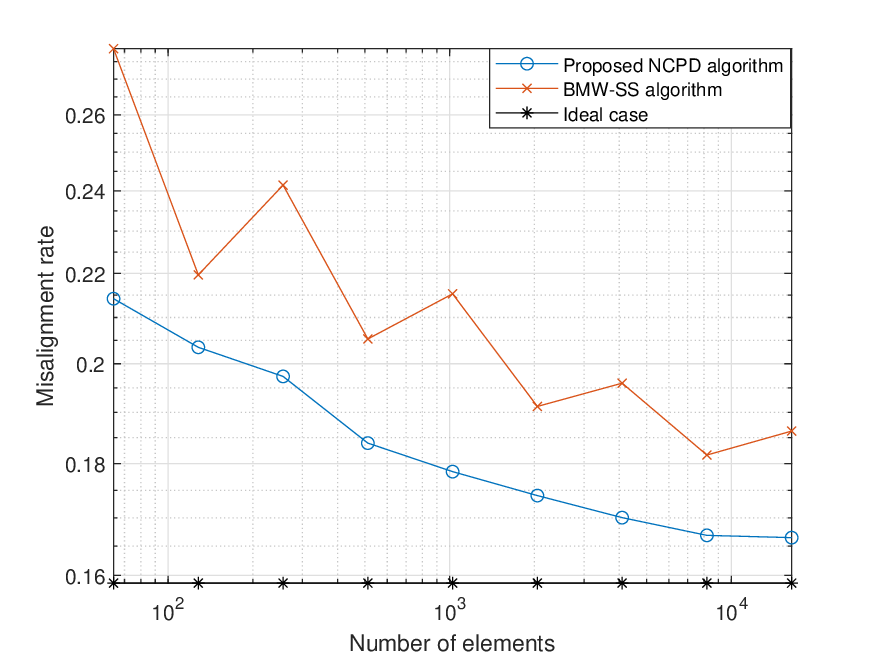}
		\caption{  The misalignment rates on different IRS size when IRS received SNR is 10 dB. }  \label{f6} 	
	\end{figure}
	
	Fig. \ref{f5} and Fig. \ref{f6} show the misalignment rates of the NCPD  and  BMW-SS algorithms on ULAs of different sizes when the IRS received SNR is 10dB and 20dB, respectively. The reason why the curve of the BMW-SS algorithm is a broken line is that  the relationship between the number of sub-arrays and  number of array elements is not a continuous function. For example, when the number of elements is 128 or 256, the BMW-SS algorithm divides  IRS into 16 sub-arrays. But when the number of elements is 512, it divides  IRS into 32 sub-arrays. When the IRS array size is in the range from 128 to 256, the number of sub-arrays does not change. But due to the increase in the total number of elements, the number of elements in each sub-array increases, which leads to a more serious deep depression problem in the combined beam and an increase in the misalignment rate of the beam. This also shows that the deep depression problem is the performance bottleneck when dealing with large-scale array problems using the conventional beam combination algorithm. The performance of the NCPD algorithm is gradually improved with the increase in the array size. In all the experimental array sizes, the misalignment of the NCPD algorithm is lower than that of the BMW-SS algorithm, which proves that the NCPD algorithm is superior in solving the beamforming problem of large-scale IRS arrays.

	\section{Conclusion}
	
	Aiming at the characteristics of IRS-aided communication scenarios, in this paper we proposed a new array factor model. In this model, we proposed the variables of the cascaded channel direction  to replace the conventional elevation and azimuth angles for describing the AoA and AoD of  IRS which can reduce the number of parameters. In addition, we used the integral form to replace the summation form in the conventional array factor. 
	In order to  control the shape of the array factor function  of IRS, we proposed the NCPD algorithm and an analytical solution of beamforming on IRS.  The  time complexity of the analytical solution  was $\mathcal{O}(1)$, which solved the problem that beamforming design takes too much time on large-scale arrays. 
	Using beams formed by the analytical solution, we proposed a hierarchical codebook and two beam training schemes,  namely the  JS scheme and  DWS scheme.  These two schemes could also be used in combination to form a more flexible beam training scheme for adapting to different scenarios. 
	
	The simulation results showed that the performance of the NCPD algorithm was better than those of the conventional  beam combination algorithms  under different array sizes, and the beam formed by the NCPD algorithm did not suffer from any deep depression  problem. At the same time, the misalignment  rate of the beam formed by the NCPD algorithm was lower than those of the beam combination algorithms under different array sizes and IRS received SNR. It could be seen that the NCPD algorithm could significantly reduce the misalignment rate of the beam.
	
	There are many important issues  not yet addressed in this paper, some of which are listed below for motivating future works.
	Firstly, we assumed that the phase of each element of IRS can be configured to  any value. However, in practice, it is often not satisfied. 
	Secondly, The scenario used in this paper was a single IRS and single UE scenario. In actual situations, multiple IRSs are often deployed together.
	Thirdly, this paper focused on scenarios of static UE. In a UE movement scenario, how to balance the beam training time and beamforming gain of IRS is worthy of further investigation.
	Finally, in this paper we considered the scenario where both  BS and  UE are in the far field of  IRS. However,  IRS often needs to be deployed close to one of  BS and UE. Therefore, the near-field scenario  is more common.

	% or
	

	\appendices
	
	\section{Proof of Proposition 1}\label{appendixB}
	Suppose that there are two directions $\beta_{{\rm hor},1}$ and $\beta_{{\rm hor},2}$ satisfying:
	\begin{equation}\label{b1b2equation}
		kd\beta_{{\rm hor},1}=kd\beta_{{\rm hor},2}+2m'\pi,
	\end{equation}
	where $m'$ is an integer. For any $g_{\rm hor}(n_{\rm hor})$,   the following equation holds:
	\begin{equation}\label{P1=P2}
		\begin{array}{l}
			AFM_{\rm norm}{\beta _{{\rm hor},2}}) = \Bigg|\sum\limits_{{n_{\rm hor}} = 0}^{n_{\rm hor} - 1} {{e^{{\jmath}kd[{\beta _{{\rm hor},2}}{n_{\rm hor}} + {g_{\rm hor}}({n_{\rm hor}})]}}} \Bigg|\\
			= \Bigg|\sum\limits_{{n_{\rm hor}} = 0}^{N_{\rm hor} - 1} {{e^{{\jmath}kd[{\beta _{{\rm hor},1}}{n_{\rm hor}} + {g_{\rm hor}}({n_{\rm hor}})] + j2m\pi }}\Bigg|} \\
			= \Bigg|\sum\limits_{{n_{\rm hor}} = 0}^{n_{\rm hor} - 1} {{e^{{\jmath}kd[{\beta _{{\rm hor},1}}{n_{\rm hor}} + {g_{\rm hor}}({n_{\rm hor}})]}}}\cdot {e^{j2m\pi }}\Bigg|\\
			= \Bigg|\sum\limits_{{n_{\rm hor}} = 0}^{n_{\rm hor} - 1} {{e^{{\jmath}kd[{\beta _{{\rm hor},1}}{n_{\rm hor}} + {g_{\rm hor}}({n_{\rm hor}})]}}} \Bigg|\\
			= AFM_{\rm norm}({\beta _{{\rm hor},1}}).
		\end{array}
	\end{equation}
	From the (\ref{P1=P2}) we can see that no matter what value $g_{\rm hor}(n_{\rm hor})$ takes, $AFM_{\rm norm}({\beta _{{\rm hor},1}})$ is always equal to $AFM_{\rm norm}({\beta _{{\rm hor},2}})$.  This means that if the IRS has a beam in the ${\beta _{{\rm hor},1}}$ direction, there must also be a beam in the ${\beta _{{\rm hor},2}}$ direction, which will cause interference to other users.  We need to choose suitable $d$ to make sure there does not exist $\beta_{{\rm hor},1}$ and $\beta_{{\rm hor},2}$ satisfy (\ref{b1b2equation}).  From (\ref{betah}) and (\ref{betav}) we know the range of ${\beta _{\rm hor}}$ is $(-2,2)$, so $kd\leq \frac{\pi}{2}$, $d \leq \frac{\lambda}{4}$. Thus, the max element spacing distance on IRS is $\frac{\lambda}{4}$. Without loss of generality, we choose $d=\frac{\lambda}{4}$ in this paper.
	
	\begin{remark}
		The conclusion of the maximum element spacing distance  on the IRS is different from that on the active array. That is because beamforming on IRS serves cascaded channel while beamforming on active arrays serve one hop channel. For active array the range of $\beta_{\rm hor}$ is $(-1,1)$ while on IRS it is $(-2,2)$, so the max element spacing distance on active array is $\frac{\lambda}{2}$ while on IRS is $\frac{\lambda}{4}$.
	\end{remark}

	\section{Proof of Proposition 2}\label{appendixA}
	Since we need to find a general solution to the beamforming problem for arbitrary array sizes, we need to replace $g_{\rm hor}(n_{\rm hor})$ with a continuous function. We let:
	\begin{equation}
		{g_{\rm hor}}({n_{\rm hor}}) = \frac{1}{d}\int_0^{{n_{\rm hor}}d} { - f(u){\rm d}u},
	\end{equation}
	where $f(u)$ is a continuously differentiable function.  So (\ref{u array factor}) can be written as:
	\begin{equation}\label{Pwithcontinuesf}
		\begin{array}{l}
			AFM_{\rm norm}({\beta _{\rm hor}}) =\frac{1}{N_{\rm hor}} \Bigg|\sum\limits_{{n_{\rm hor}} = 0}^{N_{\rm hor} - 1} {{e^{{\jmath}k[d{\beta _{\rm hor}}{n_{\rm hor}} - \int_0^{{n_{\rm hor}}d} {f(u){\rm d}u} ]}}} \Bigg|\\
			=\frac{1}{N_{\rm hor}d} \Bigg|\sum\limits_{{n_{\rm hor}} = 0}^{N_{\rm hor} - 1} {{e^{{\jmath}k[d{\beta _{\rm hor}}{n_{\rm hor}} - \int_0^{{n_{\rm hor}}d} {f(u){\rm d}u} ]}}} d \Bigg|.
		\end{array}
	\end{equation}
	Let $\dot{u}=n_{\rm hor}d$. If the IRS array size tends to be infinite while IRS physical size is fixed, which means that $N_{\rm hor}d$ is a constant and $N$ tends to be infinite, so $d$ tends to $0$ and  (\ref{Pwithcontinuesf}) can be written as a integral form:
	\begin{equation}\label{appendixA3}
		AFM_{\rm norm}({\beta _{\rm hor}}) =\frac{1}{N_{\rm hor}d} \Bigg|\int_0^{N_{\rm hor}d} {{e^{{\jmath}k[{\beta _{\rm hor}}\dot{u} - \int_0^{\dot{u}} {f(u){\rm d}u]} }}} {\rm d}\dot{u}\Bigg|.
	\end{equation}
	Let $\ddot{u}=\frac{\dot{u}}{N_{\rm hor}d}$. Then, (\ref{appendixA3}) can be written as:
	\begin{equation}\label{continuous array}
		AFM_{\rm norm}({\beta _{\rm hor}}) = \Bigg|\int_0^{1} {{e^{{\jmath}k[{\beta _{\rm hor}}N_{\rm hor}d\ddot{u} - \int_0^{\ddot{u}N_{\rm hor}d} {f(u){\rm d}u]} }}} {\rm d}\ddot{u}\Bigg|.
	\end{equation}	
	Let $\mu=\frac{u}{N_{\rm hor}d}$. We further have:
	\begin{equation}\label{muQd}
		\int_0^{\ddot{u}N_{\rm hor}d} {f(u){\rm d}u}=N_{\rm hor}d \int_0^{\ddot{u}} {f[J(\mu)]{\rm d}{\mu}},
	\end{equation}
	where $J(\mu)=N_{\rm hor}d\mu$.
	Note that the (\ref{muQd}) is an integral with respect to $\mu$. By setting $\dot{f}(\cdot)=f[J(\cdot)]$, we rewrite (\ref{muQd}) as:
	\begin{equation}\label{integral of mu}
		\int_0^{\ddot{u}N_{\rm hor}d} {f(u){\rm d}u}=N_{\rm hor}d \int_0^{\ddot{u}} {\dot{f}(\mu){\rm d}{\mu}}.
	\end{equation}   
	Substituting (\ref{integral of mu}) into (\ref{appendixA3}), we can get:  
	\begin{equation}\label{appendixA6}
		AFM_{\rm norm}({\beta _{\rm hor}}) = \Bigg|\int_0^1 {{e^{{\jmath}kN_{\rm hor}d[{\beta _{\rm hor}}\ddot{u}  - \int_0^{\ddot{u}} {\dot{f}(\mu){\rm d}{\mu}} ]}}} {\rm d}{\ddot{u}} \Bigg|.
	\end{equation}
	From (\ref{appendixA6}), we can get that the IRS array factor can be determined by a function $\dot{f}(\mu)$ defined on $(0,1)$. Thus (\ref{g_represent}) holds. Since $\ddot{u}=\frac{n_{\rm hor}}{N_{\rm hor}}$, the relationship between $g_{\rm hor}(n_{\rm hor})$ and $\dot{f}(\mu)$ is:
	\begin{equation}\label{g&fmu}
		{g_{\rm hor}}({n_{\rm hor}}) = -N_{\rm hor}\int_0^{\frac{{{n_{\rm hor}}}}{N_{\rm hor}}} {\dot{f}(\mu )d\mu }.
	\end{equation}
	

	\ifCLASSOPTIONcaptionsoff
	\newpage
	\fi

	\bibliographystyle{IEEEtran} 
	\bibliography{myref}  
\end{document}